\documentclass{aa} 

\usepackage{graphicx}
\usepackage{url}
\usepackage[breaklinks=true]{hyperref}
\usepackage{twoopt}
\usepackage{natbib}
\bibpunct{(}{)}{;}{a}{}{,} %% natbib format for A&A and ApJ
\usepackage{amssymb}
\usepackage{float}
\usepackage{color}
\usepackage{epstopdf}
\usepackage{ctable}
\usepackage{multirow}

\newcommand{\adeg}[1]{{#1}$^{\circ}$}
\newcommand{\amin}[1]{{#1}$^\prime$}
\newcommand{\asec}[1]{{#1}$^{\prime\prime}$}
\newcommand{\thour}[1]{{#1}$^{\mathrm{h}}$}
\newcommand{\tmin}[1]{{#1}$^{\mathrm{m}}$}
\newcommand{\tsec}[1]{{#1}$^{\mathrm{s}}$}

\newcommand{\hms}[3]{\thour{#1}\tmin{#2}\tsec{#3}}
\newcommand{\dms}[3]{\adeg{#1}\amin{#2}\asec{#3}}

\begin{document}

\title{Revived Fossil Plasma Sources in Galaxy Clusters}
\titlerunning{Revived Fossil Plasma Sources in Galaxy Clusters}

\author{S.~Mandal\thanks{E-mail: mandal@strw.leidenuniv.nl}\inst{1}\and H.~T.~Intema\inst{1,2}\and R.~J.~van~Weeren\inst{1}\and T.~W.~Shimwell\inst{1,3} \and A.~Botteon\inst{1,4,5} \and G.~Brunetti\inst{4} \and F.~de~Gasperin\inst{6} \and M.~Br\"uggen\inst{6} \and G.~Di Gennaro\inst{1}\and R.~Kraft\inst{7}\and H.~J.~A.~R\"ottgering\inst{1} \and M.~Hardcastle\inst{8}\and C.Tasse\inst{9,10}}

\institute{Leiden Observatory, Leiden University, PO Box 9513, NL-2300 RA Leiden, The Netherlands \and
International Centre for Radio Astronomy Research -- Curtin University, GPO Box U1987, Perth,
WA 6845, Australia \and
ASTRON, the Netherlands Institute for Radio Astronomy, Postbus 2, NL-7990 AA Dwingeloo, The Netherlands \and
INAF - IRA, via P.~Gobetti 101, I-40129 Bologna, Italy \and
Dipartimento di Fisica e Astronomia, Universit\`{a} di Bologna, via P.~Gobetti 93/2, I-40129 Bologna, Italy \and 
Hamburger Sternwarte, Gojenbergsweg 112, D-21029 Hamburg, Germany \and
Smithsonian Astrophysical Observatory, 60 Garden St., MS-4, Cambridge, MA 02138 \and 
Centre for Astrophysics Research, School of Physics, Astronomy and Mathematics, University of Hertfordshire, College Lane, Hatfield AL10 9AB, UK \and
GEPI \& USN, Observatoire de Paris, Université PSL, CNRS, 5 Place Jules Janssen, 92190 Meudon, France \and 
Department of Physics \& Electronics, Rhodes University, PO Box 94, Grahamstown, 6140, South Africa}
\authorrunning{S.~Mandal et al.}

\date{\today}
% These dates will be filled out by the publisher
\date{Accepted XXX. Received YYY; in original form ZZZ}

% Enter the current year, for the copyright statements etc.
%\pubyear{2017}

\label{firstpage}

\abstract{It is well established that particle acceleration by shocks and turbulence in the intra-cluster medium can produce cluster-scale synchrotron emitting sources. However, the detailed physics of these particle acceleration processes is still not well understood. One of the main open questions is the role of fossil relativistic electrons that have been deposited in the intra-cluster medium by radio galaxies. These synchrotron-emitting electrons are very difficult to study, as their radiative life time is only tens of Myrs at GHz frequencies, and are therefore a relatively unexplored population. Despite the typical steep radio spectrum due to synchrotron losses, these fossil electrons are barely visible even at radio frequencies well below a GHz. However, when a pocket of fossil radio plasma is compressed, it boosts the visibility at sub-GHz frequencies, creating so-called radio phoenices. This compression can be the result of bulk motion and shocks in the ICM due to merger activity. 
In this paper, we demonstrate the discovery potential of low frequency radio sky surveys to find and study revived fossil plasma sources in galaxy clusters. We used the 150~MHz TGSS {\color{black}{and 1.4 GHz NVSS}} sky surveys to identify candidate radio phoenices. A subset of three candidates were studied in detail using deep multi-band radio observations (LOFAR and GMRT), X-ray (\textit{Chandra} or \textit{XMM-Newton}) and archival optical observations. Two of the three sources are new discoveries. Using these observations, we identified common observational properties (radio morphology, ultra-steep spectrum, X-ray luminosity, dynamical state) that will enable us to identify this class of sources more easily, and helps to understand the physical origin of these sources.}

\keywords{
Radio Astronomy -- shock waves -- X-rays: galaxies: clusters -- galaxies: clusters: individual: Abell~2593, Abell~2048, SDSS-C4-DR3-3088 -- radio continuum: general -- radiation mechanisms: non-thermal
}

\maketitle

%===============================================================================

\section{Introduction}

Galaxy clusters are the largest gravitationally bound systems in the Universe. Their main mode of growth is through accretion of smaller groups of galaxies, and through major mergers with other clusters. A huge amount of gravitational binding energy of the order of $\sim 10^{64}$~ergs is released when galaxy clusters merge (e.g. \citealt{kravstov12}). This energy strongly affects the physical properties of the different constituents of clusters, e.g., the density distribution and velocity dispersion of the galaxies, and the temperature, metallicity and density distribution of the (X-ray-emitting) thermal intra-cluster medium (ICM). Cluster mergers also generate shocks which cause density and temperature jumps (\citealt{markevitch07}) and turbulence in the ICM, which can amplify magnetic fields ($\sim\mu$G), affect the spatial transport and (re)-accelerate relativistic particles (Lorentz factor $\gamma > 1000$). This particle acceleration can lead to cluster-scale synchrotron emission most commonly in the form of central radio halos and peripheral radio relics (e.g., see \citealt{vanweeren19} for recent observational reviews, and \citealt{brunettijones14} for a theoretical review). The spectral index ($\alpha$)\footnote{$S\propto\nu^{\alpha}$, where $S$ is the measured flux density and $\nu$ is the observed frequency} of this emission is generally steep ($\alpha \lesssim -1$), making them relatively bright at radio frequencies below a few hundred MHz. 

\ctable[topcap,center,star,%sideways,
caption = {Target list of 3 TGSS clusters with their properties as given in the literature.},
label = tab:sample,
doinside=\small,
]{c c c c c c c}
{
%\tnote[a]{X-ray luminosities are computed from the ROSAT count rates reported in Boller+ (2016). There is no entry of any source close to the position of this cluster.}
}
{
\FL Cluster name / Target & RA & Dec & Redshift & Arcsec to kpc conversion     
\NN                        & (J2000)            & (J2000)           & (z)       &      
\ML Abell~2593             & \hms{23}{24}{13.0} & \dms{+14}{41}{38} & 0.041  & 0.811  
\NN SDSS-C4-DR3-3088       & \hms{09}{46}{45.4} & \dms{+54}{25}{37} & 0.046  & 0.904        
\NN Abell~2048             & \hms{15}{15}{09.7} & \dms{+04}{24}{37} & 0.097 & 1.797
\LL
}

Radio halos are extended ($\gtrsim 1$~Mpc) diffuse radio sources at the center of merging clusters (e.g., \citealt{cassano10a}; \citealt{cuciti15}; \citealt{kale15}), with a morphology similar to the X-ray morphology of the system. Radio halos have been discovered in $\sim$50 disturbed clusters (e.g., \citealt{vanweeren19}). In the last decade, observational evidence suggests that they are caused by the continuous acceleration of electrons in turbulent gas (\citealt{brunetti01}; \citealt{petrosian01}, \citealt{brunettilazarian07}). In addition, continuous injection of secondary particles through proton-proton collisions (\citealt{dennison80}, \citealt{ensslin11}) can also be re-accelerated together with the primary electrons (\citealt{brunettilarazin11}, \citealt{pinzke17}) in these regions and may play a role in the generation of radio halos. However, lack of gamma ray detection puts constraints on such a mechanism (e.g., \citealt{brunettizimmer17} and references therein). 

Apart from radio halos, diffuse radio sources in clusters can broadly be divided into radio relics (e.g., \citealt{kempner04}; \citealt{vanweeren19}) and radio phoenices (revived fossil plasma sources). Cluster radio relics are elongated linearly polarized (>10\%-30\% at GHz frequencies) regions ($\sim$ 1-2~Mpc) at the cluster outskirts, which typically have a convex morphology with respect to the cluster centre and exhibit a radio spectral steepening towards the cluster centre (\citealt{vanweeren10}; \citealt{bonafede12}; \citealt{stroe13}; \citealt{gasperin15}; \citealt{kierdorf17}). Radio relics are generally found to trace merger-induced shock waves (e.g., \citealt{finoguenov10}; \citealt{bourdin13}; \citealt{akamatsu13}; \citealt{shimwell15}; \citealt{botteon16}; \citealt{eckert16a}; \citealt{urdampilleta18}). The particle (re)-acceleration processes at the location of radio relics is still debated. There are cases where the luminosity of the radio relics is much higher than expected from diffuse shock acceleration (DSA) of electrons in the thermal pool (e.g., \citealt{brunettijones14}, \citealt{botteon16}, \citealt{vanweeren16a}, \citealt{botteon19}). Therefore, the role of fossil plasma (seed particles) in the ICM has been invoked, as simulations indicate that (re)-acceleration of seed relativistic electrons is more efficient than the DSA (e.g., \citealt{kangryu15}). 

Radio phoenices are a less widely studied class of diffuse radio sources in the ICM, which are thought to be a manifestation of fossil plasma in galaxy clusters. Likely candidates for the fossil electrons are lobes and tails of radio galaxies, which have ultra-steep spectra (USS) due to synchrotron aging and Inverse Compton (IC) losses. After being deposited in the ICM, these fossil electrons are visible only for tens of Myrs at GHz frequencies. When an aged lobe or tail is compressed (e.g., by a merger shock), the electrons are re-energised which boosts their visibility at sub-GHz frequencies (\citealt{eg01}; \citealt{ensslinbruggen02}). Sources powered by this mechanism can maintain electrons at higher energies than what radiative cooling alone would allow. Although deposited locally, when given enough time, fossil plasma can occupy a significant fraction of the (turbulent) ICM due to advection and diffusion. The relative importance of this mechanism to explain the origin of the other diffuse cluster radio sources (halos, relics) is unclear. To date, only a few radio phoenices have been found using low-frequency observations of clusters (e.g. \citealt{slee01}; \citealt{kempner04}; \citealt{vanweeren09}; \citealt{vanweeren11}; \citealt{gasperin15}, \citealt{mandal18}). Even though radio tails and lobes are abundant in clusters, the ultra-steep nature of these sources makes it hard to identify, especially from large-area sky surveys at GHz frequencies. To firmly establish radio phoenices as a distinct class of objects, it is crucial to identify their common observable and physical properties, and compare these with radio halos and relics. Identifying common observational properties also enables an efficient search for more radio phoenices, which in itself will help to better understand the nature and physical mechanisms behind this relatively unexplored population of radio sources.

Radio sky surveys below 300~MHz such as the VLA Low-frequency Sky Survey (VLSS; \citealt{cohen07}; \citealt{lane12}; \citealt{lane14}), the GaLactic and Extragalactic All-sky Murchison Wide-field Array Survey (GLEAM; \citealt{wayth15}, \citealt{hurley-walker17}), and the LOFAR Multi-frequency Snapshot Sky Survey (MSSS; \citealt{heald15}) have a great potential for discovery in this area, but the low spatial resolution of these surveys makes it difficult to distinguish between diffuse radio emission and emission from individual, active radio galaxies. The TIFR GMRT Sky Survey (TGSS ADR; \citealt{intema17}) is a high-resolution sky survey at 150~MHz. The combination of very large area (about 37,000 square degrees, all radio sky between \adeg{-53} to \adeg{+90}), high resolution (\asec{25}) and high sensitivity (2-5 mJy beam$^{-1}$ rms noise) makes the TGSS a very suitable resource to search for radio phoenices based on their diffuse morphology, ultra-steep radio spectrum, and location in galaxy clusters. 

In this paper we present a pilot study for identifying the properties of radio phoenices and their interplay with the dynamics and properties of the hosting clusters from multi-band radio observations, complemented with optical and X-ray observations. We selected 11~radio phoenix candidates from the TGSS based on their morphology, steep spectra, and location (see Section \ref{section:sample}). {\color{black}{For three of the candidates, we managed to get dataset at three different radio (LOFAR 150 MHz, GMRT 325 MHz and 610 MHz) and X-ray (\textit{Chandra} or \textit{XMM-Newton}) band which we are presenting in this paper as an exploratory study.}} The layout of the paper is as follows: In Section~\ref{section:sample} we present the method of the sample selection. In Section~\ref{section:obs} we describe the observations and data reduction methods. In Section~\ref{section:results} we discuss the results deduced from the radio, X-ray, optical and spectral index images of these sources, followed by discussion in Section~\ref{section:discussion} and conclusions in Section~\ref{section:conclusions}.

Throughout the paper, we assume a $\Lambda$CDM cosmology with $H_0$ = 70 kms$^{-1}$Mpc$^{-1}$, $\Omega_m$ = 0.3, and $\Omega_\Lambda$ = 0.7. All sky coordinates are epoch J2000 coordinates.

\ctable[botcap,center,star,
caption = {Observation Details (on-source time) of Abell~2593, SDSS-C4-DR3-3088 and Abell~2048.},
label = tab:obs-xray
]{l c c c c c}{
%\tnote[a]{Specified for the final (second) calibration cycle only.} \tmark[a]
}{
\FL {Source} & LOFAR 150 MHz & GMRT 325 MHz & GMRT610 MHz & \textit{Chandra} & \textit{XMM-Newton} 
\NN {} & (120-168)MHz & (308-340)MHz & (594-626)MHz & (0.5-2.0)keV &(0.5-2.0)keV
\ML Abell~2593 & 8.33 h & 3.83 h & 3.18 h & 7 ks & NA 
\NN SDSS-C4-DR3-3088 & 8.33 h & 3.68 h & 3.55 h & 17 ks & NA
\NN Abell~2048 & 8.33 h & 10 h & 10 h & NA & 68 ks
\LL
}

\begin{table*}[h!]
\centering
\resizebox{0.65\textwidth}{!}{
\begin{tabular}{llllll}
Soure Name       & Frequency     & Robust & Taper & $\Theta_{FWHM}$  & $\sigma_{rms}$ \\
                 & (MHz)         &        & (\asec{}) & \asec{} $\times$ \asec{}, (PA) & $\mu$Jy beam$^{-1}$
\\
\\
\hline
Abell~2593       & LOFAR 150 & -0.5   & 0         & 9$\times$6, (75.45)         & 220  \\
                 & GMRT 325  & -0.5   & 0         & 11$\times$7, (18.53)        & 66   \\
                 & GMRT 610  & -0.5   & 0         & 5$\times$4, (20.75)         & 38   \\
\hline
SDSS-C4-DR3-3088       & LOFAR 150 & -0.25   & 0         & 6$\times$6, (90.00)         & 80  \\
                       & GMRT 325  & -0.25   & 0         & 13$\times$7, (46.18)        & 43   \\
                       & GMRT 610  & -0.25   & 0         & 7$\times$4, (-21.82)         & 40   \\
\hline
Abell~2048       & LOFAR 150 & -0.5   & 0         & 13$\times$6, (85.74)         & 266  \\
                 & GMRT 325  & -0.5   & 0         & 11$\times$8, (66.60)        & 125   \\
                 & GMRT 610  & -0.5   & 0         & 8$\times$5, (-63.51)         & 117   \\

\hline
\end{tabular}
}
\caption{Imaging parameters and image properties of the LOFAR and GMRT maps of Abell~2593, SDSS-C4-DR3-3088 and Abell~2048 respectively. The beam position angle (PA) is measured in degrees from north through east.}
\label{tab:image-parameter}
\end{table*}

\ctable[botcap,center,star,
caption = {Global properties of the clusters derived from X-ray observations. Quoted errors are at the 1$\sigma$ confidence level. $R_{500}$ and $M_{500}$ are derived using the scaling relations from \citealt{arnaud05}.
},
label = tab:xray-parameter
]{l c c c c c c c}{

}{
\FL Cluster &  <kT> & Metallicity (Z) & $R_{500}$ & $M_{500}$
\NN     & (keV) & ($Z_\odot$) & (kpc) & $M_\odot$ $\times$ $10^{14}$ 
\ML Abell~2593 & 4.16 $\pm$ 0.28 & $0.28^{+0.19}_{-0.17}$ & $1011^{+33}_{-35}$ & $3.06^{+0.31}_{-0.30}$ 
\NN SDSS-C4-DR3-3088  & 3.13 $\pm$ 0.41 & $0.79^{+0.53}_{-0.37}$ & $829^{+59}_{-65}$ & $1.69^{+0.39}_{-0.36}$
\NN Abell~2048  & 4.15 $\pm$ 0.10 & $0.55^{+0.06}_{-0.06}$ & $982^{+12}_{-11}$ & $2.97^{+0.10}_{-0.11}$  
\LL}

%===============================================================================

\section{Sample Selection}
\label{section:sample}

To search for candidate radio phoenices in galaxy clusters, we convolved the TGSS images (resolution \asec{25}) to the resolution of NVSS (The NRAO VLA Sky Survey; \citealt{condon98}: resolution \asec{45}) images, combined the two, and identified extended, steep-spectrum emission. We selected sources based on the following properties: 
\begin{itemize}
\item{An ultra-steep radio spectrum ($\alpha \leq -2$)}
\item{A 150~MHz total flux density $S_{150}$ $\geq$ 30 mJy (detection above $\sim$ 10$\sigma$)}
\item{A morphology compatible with diffuse cluster emission (no point sources associated with dead AGNs)}
\item{Located in the vicinity (within 1~Mpc) of a known cluster position from the SDSS cluster catalogue by \citealt{wen12} or the ROSAT All-Sky Survey (RASS; \citealt{rass99}) catalogue}
\end{itemize}
The limiting values of the spectral index and flux density in selecting the sample is somewhat arbitrary, aimed at identifying very steep-spectrum, extended and relatively bright candidates for enabling a more detailed study using follow-up observations. This bright sample is likely the tip of the iceberg of the population and in this way, the contamination of other types of sources (such as radio relics / halos) is also minimal. Our final sample consists of 11~candidate revived fossil plasma sources. For this study we performed follow-up observations for three of the sources with LOFAR~150 MHz, GMRT 325~MHz, GMRT 610~MHz and \textit{Chandra} and/or \textit{XMM-Newton} X-ray observations. An overview of these three objects is given in Table~\ref{tab:sample}. The detailed observation summary is given in Table~\ref{tab:obs-xray}. The remainder of our sample will be followed up with future observations.

%===============================================================================

\section{Observations and Methods}
\label{section:obs}

\subsection{GMRT Data Reduction}

We observed two of our targets (Abell~2593 and SDSS-C4-DR3-3088) with the GMRT at 325 and 610~MHz (project codes 31\_018 and 33\_014), while for the third target (Abell 2048) we used archival observations at the same frequencies (project codes 15HRA01 and 16\_065). The data were processed using the SPAM pipeline \citep{intema17} which includes Radio Frequency Interference (RFI) mitigation schemes, direction-dependent calibration, and ionospheric modelling \citep{intema09}. The final full-resolution GMRT images are shown in the middle panels of Figures \ref{fig:a2593}, \ref{fig:sdssc4} and \ref{fig:a2048} which were created using the imaging parameters as reported in Table~\ref{tab:image-parameter}. The flux densities were set using calibration on 3C286 (at 325~MHz) and 3C48 (at 610~MHz) using the models from \cite{SH12}. We adopted a flux uncertainty of 10\% for GMRT observations (\citealt{chandra04}), which is quadratically added to the uncertainties of all flux density measurements from the GMRT observations.

\subsection{LOFAR Data Reduction}
{\color{black}{All the three sources presented in this paper were observed with the LOFAR at 150~MHz (project codes LC9\_027 for Abell~2593 \& Abell~2048 and LC6\_015 for SDSS-C4-DR3-3088) in \texttt{HBA\_DUAL\_INNER} mode}}.{\color{black}{The LOFAR data were processed with the standard direction-independent (DI) calibration pipeline \footnote{https://github.com/lofar-astron/prefactor} as described in \cite{vanweeren16a} and \cite{williams16} to correct for DI effects (see \citealt{gasperin19} for a description of the latest version of this pipeline). The processing was performed on compute facilities that are local to the LOFAR long term archive sites to mitigate issues with downloading large quantities of data to local compute clusters (see \citealt{mechev18}). After DI calibration was completed the data were processed with the latest version of the LoTSS (\citealt{shimwell17} and \citealt{shimwell19}) direction-dependent (DD) calibration and imaging pipeline that will be presented in detail in Tasse et al., (in prep.) and is also summarised in \citet[][Sec. 5.1]{shimwell19}. This pipeline makes use of kMS (\citealt{tasse14} and \citealt{smirnov15}) for DD calibration, and DDFacet \citep{tasse18} is used to apply the DD calibration solutions during the imaging.

After completion of the LoTSS data processing pipeline a post processing step is conducted to subtract all the sources besides the targets and calibrate in the directions of the targets. This post-processing procedure will be described in van Weeren et al. (in prep).

Due to normalisation issues and inaccurate beam models, LOFAR flux scales can show systematic offsets which needs to be corrected by comparing other surveys (\citealt{vanweeren16b}, \citealt{hardcastle16}). We cross-matched catalogues of LOFAR point sources near the targets with the TGSS (\citealt{intema17}). The adopted correction factor is at the order of 15\% on LOFAR flux densities that was derived from the total flux density ratio between the LOFAR and the TGSS. }}

\subsection{Spectral Index Maps and Integrated Spectrum Calculation}
\label{sec:spix_maps}

In Table~\ref{tab:image-parameter} we list the properties of the images made with the radio data. These images were used to create high-resolution spectral index maps between 150~MHz, 325~MHz and 610~MHz. In order to sample the same spatial scales at all frequencies, we remade the images using the same inner \textit{uv}-range of 200$\lambda$ for all the radio frequencies. Only for Abell~2048, an inner \textit{uv}-range of 600$\lambda$ was used to suppress ripples caused by calibration errors on shorter baselines. The \textit{uv}-taper parameter was varied for individual data sets (as reported in Table~\ref{tab:image-parameter}). The resulting images were convolved with a 2D~Gaussian to produce images with the same restoring beam size, accurately aligned in the image plane, and re-gridded onto the same pixel grid. {\color{black}{To calculate the spectral index maps, pixels only with a surface brightness > 3$\sigma$ that were detected in at least two images were used (where $\sigma$ is the local rms background noise). These masked images were then used to create spectral index maps where a power-law function was fit for each pixel. The errors in the spectral index were calculated taking into account the image noise and a flux scale uncertainty of 15\% for LOFAR and 10\% for GMRT. {\color{black}{Spectral index error maps are included in the Appendix A and B.}}}} 

\subsection{X-ray Data Reduction}

Abell~2593 and SDSS-C4-DR3-3088 were observed with \textit{Chandra} ACIS-I in VFAINT mode as part of the GTO program (Obs IDs 20780 and 20781). Abell~2048 was already observed twice with EPIC on-board of \textit{XMM-Newton} in Full Frame mode (Obs IDs 0653810601 and 0760230301). We retrieved these data sets (see Table~\ref{tab:obs-xray} for more details) and analyzed the \textit{Chandra} data with CIAO v4.11 and CALDB v4.8.2, while we made use of the ESAS implemented in SAS v16.1.0 to process the \textit{XMM-Newton} data. Observations were reduced following standard reduction procedures, including the removal of bad pixels, screening for periods affected by soft proton flares, and detection and excision of point sources prior to spectral extraction. We used the 0.5-2.0 keV energy band to produce the cluster images shown in this paper.

Spectral analysis was performed with XSPEC v12.10.0c (\citealt{arnaud96}) to derive global properties of the clusters. Given the low redshifts of Abell~2593 and SDSS-C4-DR3-3088 and their large angular extent in the ACIS-I field of view (FoV), we made use of the \textit{Chandra} blank sky field datasets scaled by the ratio of the 9.5-12 keV count rates to estimate the local background. The larger FoV of \textit{XMM-Newton} and smaller angular size of Abell~2048 allowed us to perform accurate modelling of the background components in a cluster-free region. We followed \cite{ghirardini19}, adopting their phenomenological model for the non-X-ray background and an astrophysical component composed of the combined cosmic X-ray background together with the Galactic foreground emission in the direction of the cluster. The ICM emission was modelled with an absorbed thermal plasma model PHABS*APEC, fixing the redshift of the cluster to the value given in Table~\ref{tab:sample} and the column density to the Galactic absorption value towards the target (\citealt{kalberla05}). The normalisation, temperature and metallicity (solar abundance table by \citealt{asplund09}) of the APEC model were free to vary in the fit.
 
\subsubsection*{Derivation of the Cluster Global Properties}

{\color{black}{We derived the global temperature and metallicity of the clusters from spectra extracted in the region 0.1$ \times R_{500}$ < r < 0.4$ \times R_{500}$, a choice that is commonly used in the literature to avoid the effect of the cool core (if present), and obtained values at the virial radius}}. As $R_{500}$ is not known for our clusters, we used an iterative approach to compute $<kT>$ and $R_{500}$ (see \citealt{liu18}, \citealt{mernier19}): we first fixed $R_{500}$ arbitrarily and evaluated the corresponding $<kT>$, then we used the scaling relations of \cite{arnaud05} to compute $R_{500}$ (and $M_{500}$) and compared it with the new $<kT>$ obtained until we obtained a self-consistent result. The global properties of the clusters are summarised in Table~\ref{tab:xray-parameter}. 

%===============================================================================

\section{Results}
\label{section:results}

Our three radio phoenix candidates are associated with the clusters as listed in Table~\ref{tab:sample}. A list of global cluster properties (temperature, metallicity, $R_{500}$ and $M_{500}$) as derived from the X-ray observations is provided in Table~\ref{tab:xray-parameter}. Below we first discuss the results for the three individual radio sources in their respective clusters.

\subsection{Abell~2593}
\label{section:results_abell2593}

\begin{figure*}[!htb]
\begin{center}
\resizebox{0.48\hsize}{!}{
\includegraphics[angle=0]{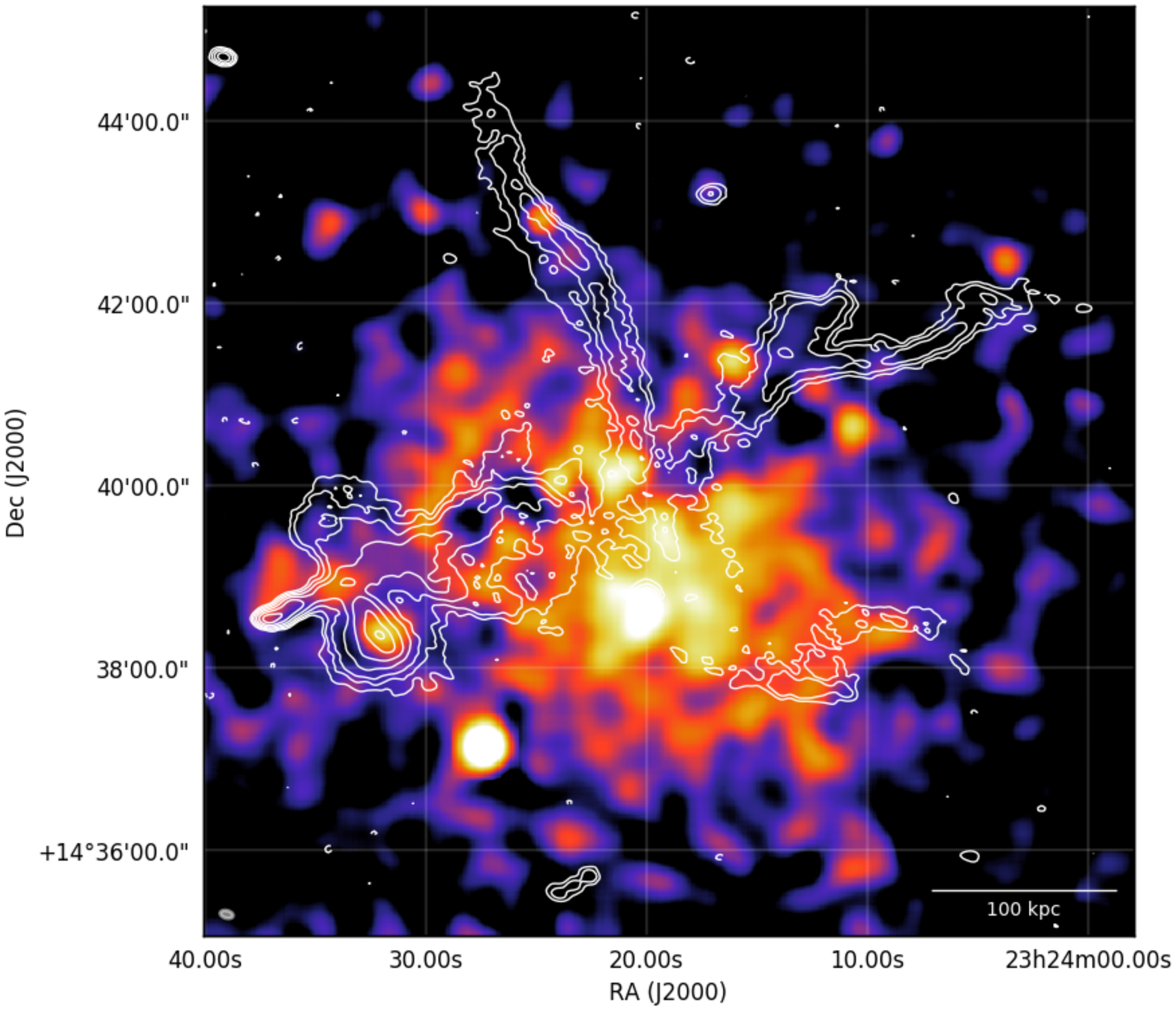}
}
\resizebox{0.49\hsize}{!}{
\includegraphics[angle=0]{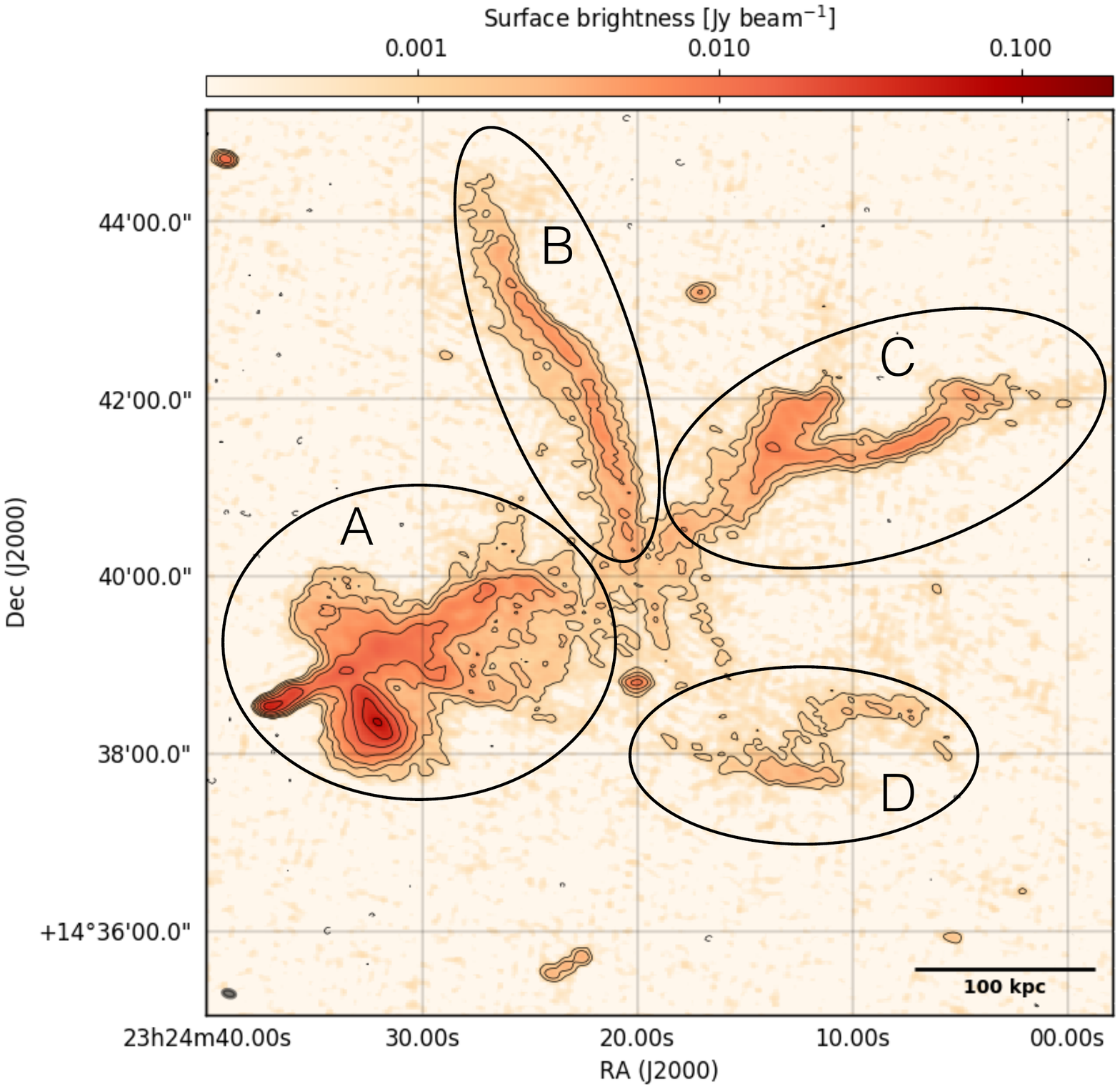}
}
\resizebox{0.49\hsize}{!}{
\includegraphics[angle=0]{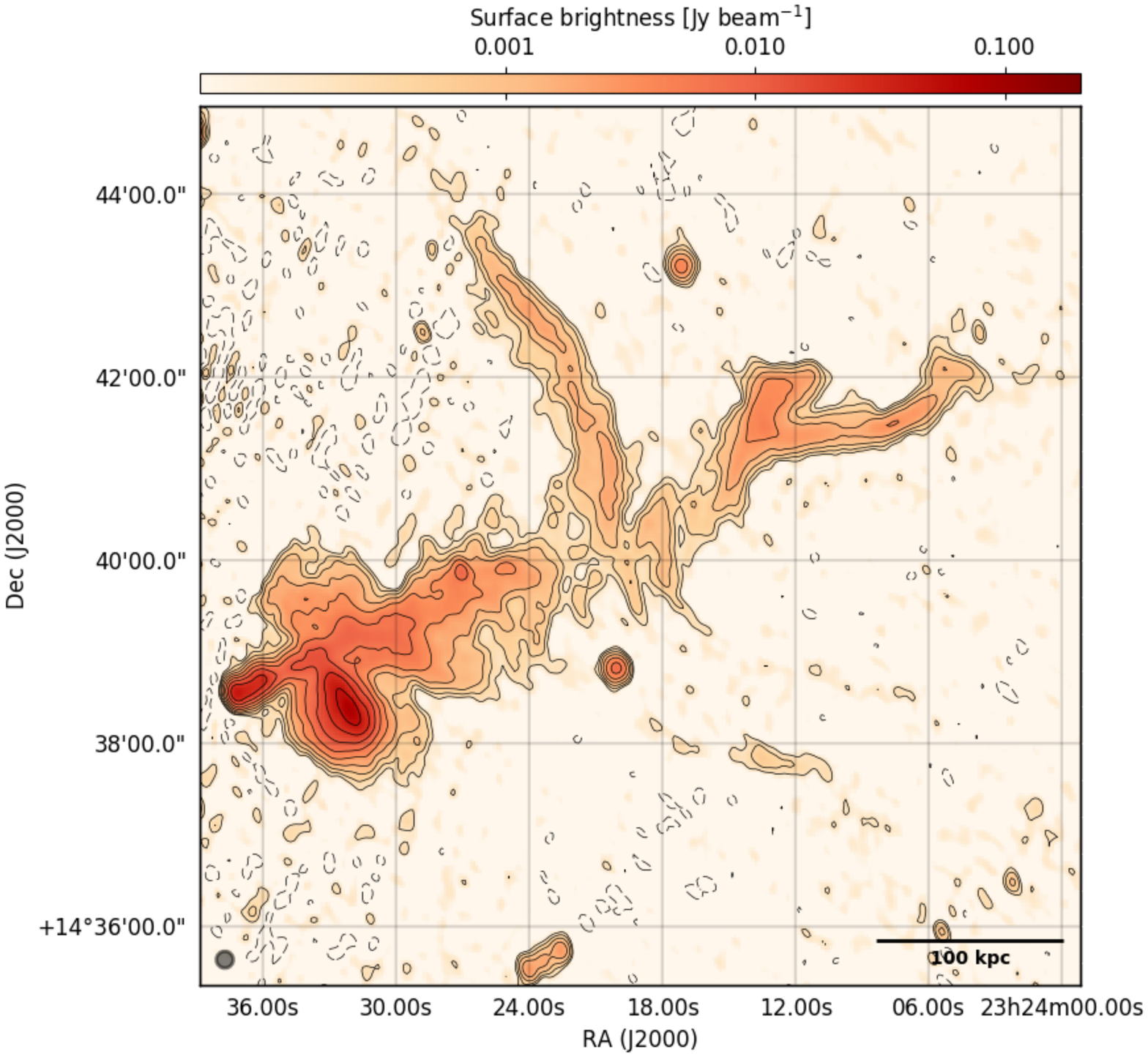}
}
\resizebox{0.49\hsize}{!}{
\includegraphics[angle=0]{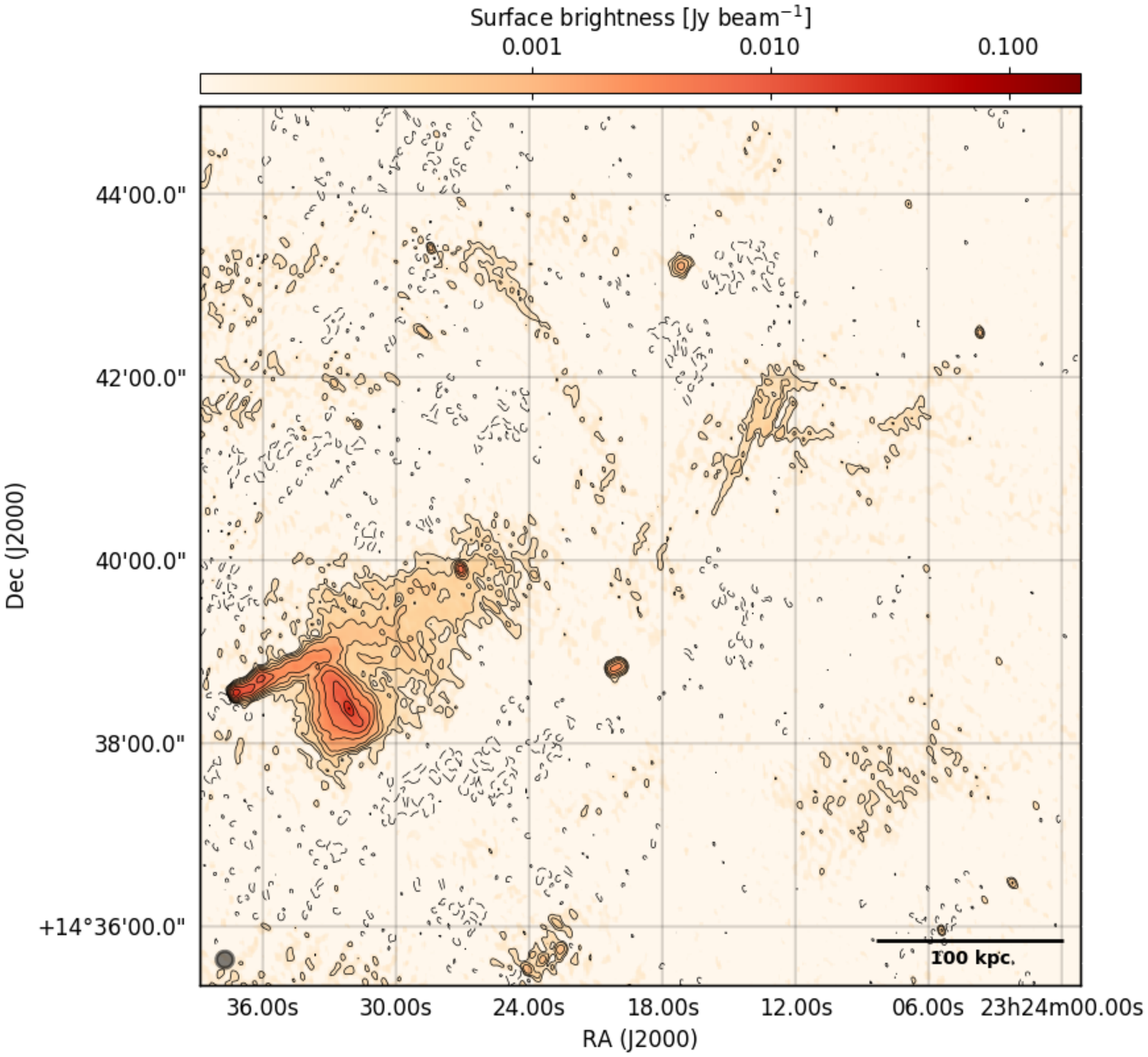}
}
\resizebox{0.49\hsize}{!}{
\includegraphics[angle=0]{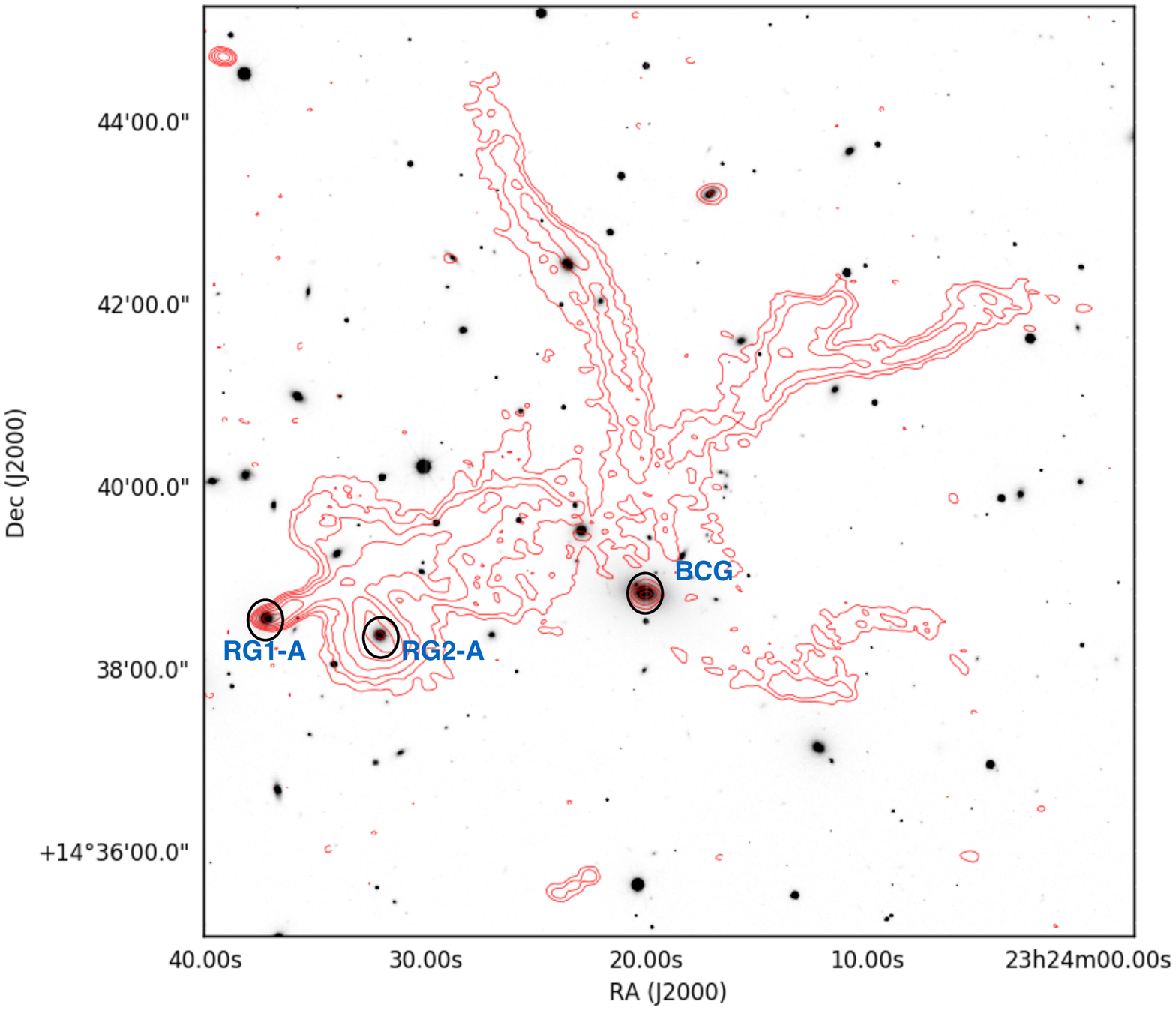}
}
\resizebox{0.48\hsize}{!}{
\includegraphics[angle=0]{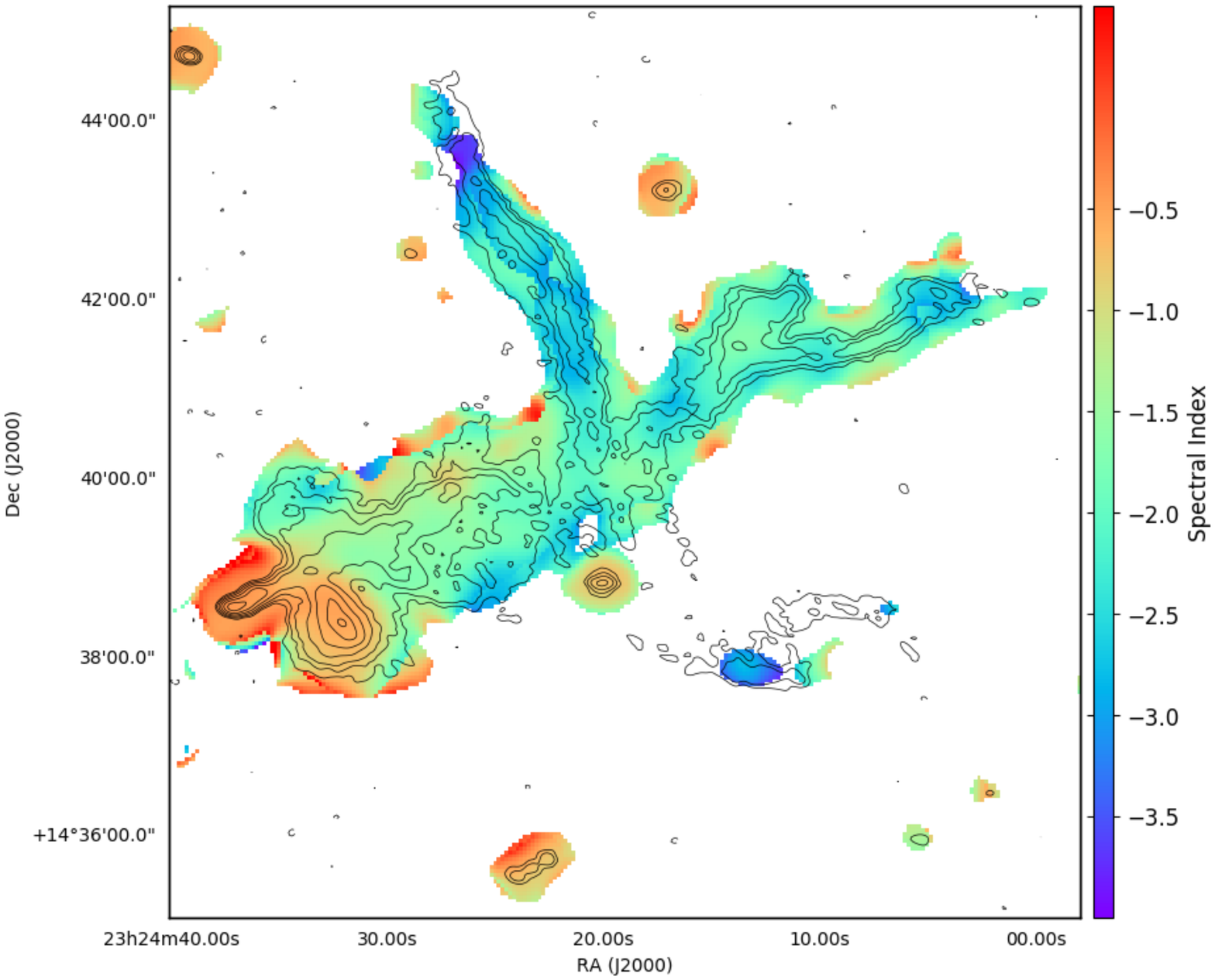}
}
\caption{
\textit{Top left} panel: Full-resolution LOFAR image contours (white; as shown in the \textit{right} panel) of Abell~2593, overlaid on an exposure-corrected \textit{Chandra} image in the 0.5-2.0 keV energy band with a total integration time of 7ks.
\textit{Top right} panel: The 150 MHz LOFAR image (\asec{9} $\times$ \asec{6}) where the black contours and dashed lines show the (1,2,4,...)$\times$5$\times$ $\sigma_{\rm{LOFAR150}}$ and -3$\times$ $\sigma_{\rm{LOFAR150}}$ levels respectively, where $\sigma_{\rm{LOFAR150}}$=219$\mu$Jy/beam. 
\textit{Middle left} panel: The 325 MHz GMRT image (\asec{11} $\times$ \asec{7}) where the black contours and dashed lines show the (1,2,4,...)$\times$5$\times$ $\sigma_{\rm{GMRT325}}$ and -3$\times$ $\sigma_{\rm{GMRT325}}$ levels respectively, where $\sigma_{\rm{GMRT325}}$=66$\mu$Jy/beam. 
\textit{Middle right} panel: The 610 MHz GMRT image (\asec{5} $\times$ \asec{4}) where the black contours and dashed lines show the (1,2,4,...)$\times$5$\times$ $\sigma_{\rm{GMRT610}}$ and -3$\times$ $\sigma_{\rm{GMRT610}}$ levels respectively, where $\sigma_{\rm{GMRT610}}$=38$\mu$Jy/beam. 
\textit{Bottom left} panel: LOFAR 150~MHz contours as shown in the \textit{top left} panel in \textit{red} overlaid on an SDSS r-band image of Abell~2593. The marked galaxies (RG1-A and RG2-A) are the possible counterparts for the radio emission of regions A and C. The BCG mark indicates the brightest cluster galaxy.   
\textit{Bottom right} panel: The LOFAR 150~MHz contours as shown in the top \textit{right} panel overplotted on the high resolution (\asec{20} $\times$ \asec{20}) spectral index map of Abell~2593. 
}
\label{fig:a2593}
\end{center}
\end{figure*} 

%----about the cluster-----%
The galaxy cluster Abell~2593 is located at a redshift of $z = 0.041$ with an Abell richness class of $R = 0$ (\citealt{ulmer81}). The top-left panel of Figure~\ref{fig:a2593} shows the image of the X-ray gas over-plotted with the contours of the 150~MHz radio emission (in white) as seen by LOFAR. The apparent morphology of the X-ray emission looks roundish and we do not see strong evidence for elongation of the emission in any particular direction, as is typically the case for merging clusters. However, the X-ray emission is not peaked at the center, suggesting that the dynamical state could be somewhat disturbed. 

%----radio morphology----%
The radio source in this cluster has a filamentary morphology with distinctive features. The top-right panel shows the high-resolution LOFAR 150~MHz image of the source that extends 450~kpc in the SE-NW direction (marked as regions~A and C), and an elongated structure extending about 200~kpc in the NE direction (marked as region~B). There is a hint of diffuse emission towards the SW (marked as region~D). The middle-left and -right panels of Figure~\ref{fig:a2593} show the GMRT 325 and 610~MHz images, respectively. The GMRT 325~MHz image has a very similar morphology to the LOFAR 150~MHz image, but there is an apparent disconnection towards the south of region~B. Region~A has a similar morphology in the GMRT 610~MHz image as well, but we do not detect the significant emission in regions~B and D. Also, the radio emission in region~C is very faint in the GMRT 610~MHz image, which is indicative of steep-spectrum radio emission. 
%----optical description----%
The bottom left panel shows the LOFAR 150~MHz contours (in red) overlaid on an SDSS r-band image (\citealt{wen12}) of Abell~2593. The brightest cluster galaxy (marked as BCG) is associated with a compact radio source visible in all three radio maps. Two other optical galaxies (marked RG1-A and RG2-A) are also found to be associated with radio emission. The peak of the X-ray emission coincides with the BCG position. {\color{black}{Given that, it is a poor cluster of galaxies (since it is of low-mass), it is hard to determine the dynamical state of the cluster based on very few galaxy cluster members.}} 

%----spectral index----%
In order to measure the integrated radio spectral index of the different regions of the radio emission Abell~2593, we made a set of images (not included in this paper) at all available radio frequencies with identical imaging parameters and minimum \textit{uv}-range (see Section \ref{sec:spix_maps} for details). The integrated flux densities of the whole source as measured over the same area (defined by the 3$\sigma$ contours of the LOFAR image) at 150~MHz, 325~MHz and 610~MHz are $S_{150} = 3.57 \pm 0.36$~Jy, $S_{325} = 1.45 \pm 0.15$~Jy, $S_{610} = 0.69 \pm 0.07$~Jy, respectively. A single power-law fit gives an integrated spectral index measurement of $\alpha_{150-325-610} =  -1.17 \pm 0.12$. This is less steep than the integrated spectral index measured between 150~MHz TGSS and 1.4~GHz NVSS flux densities (This apparent discrepancy is discussed in Section \ref{section:discussion}).
We used the LOFAR 150~MHz, GMRT 325 and 610~MHz maps to create a spatially resolved spectral index map (see Section~\ref{sec:spix_maps}) to look for trends in the variation of $\alpha$ at a resolution of \asec{20}. The bottom-right panel of Figure~\ref{fig:a2593} shows the spectral index distribution across the source. The spectral index values at the location of RG1-A and RG2-A (region~A) are -0.8 (\textit{flat}). For the source RG1-A, the spectral index steepens in the NW direction to a value of -1.7. The regions~B and C are more uniformly steep with small variations ranging from -1.7 to -2.0. Region~D has a very steep spectral index value of -3.0. We note that the compact radio emission from the BCG has a spectral index of -0.7.

{\color{black}{The integrated spectral index value of the entire source from 150~MHz--325~MHz--610~MHz is not very steep due to the contributions of the radio flux at the location of RG1-A, RG2-A, and the BCG and is not an accurate representative of the spectral index of the diffuse regions of the source.}} The spectral index value tends to get steeper when we add flux density measurements from 1.4~GHz NVSS images which suggests a possibly curved spectrum at higher frequencies. However, deeper 1.4~GHz observations are needed in order to confirm the spectral curvature. 

%----interpretation----%

\subsection{SDSS-C4-DR3-3088}
\label{section:results_sdssc4}

\begin{figure*}[!htb]
\begin{center}
\resizebox{0.49\hsize}{!}{
\includegraphics[angle=0]{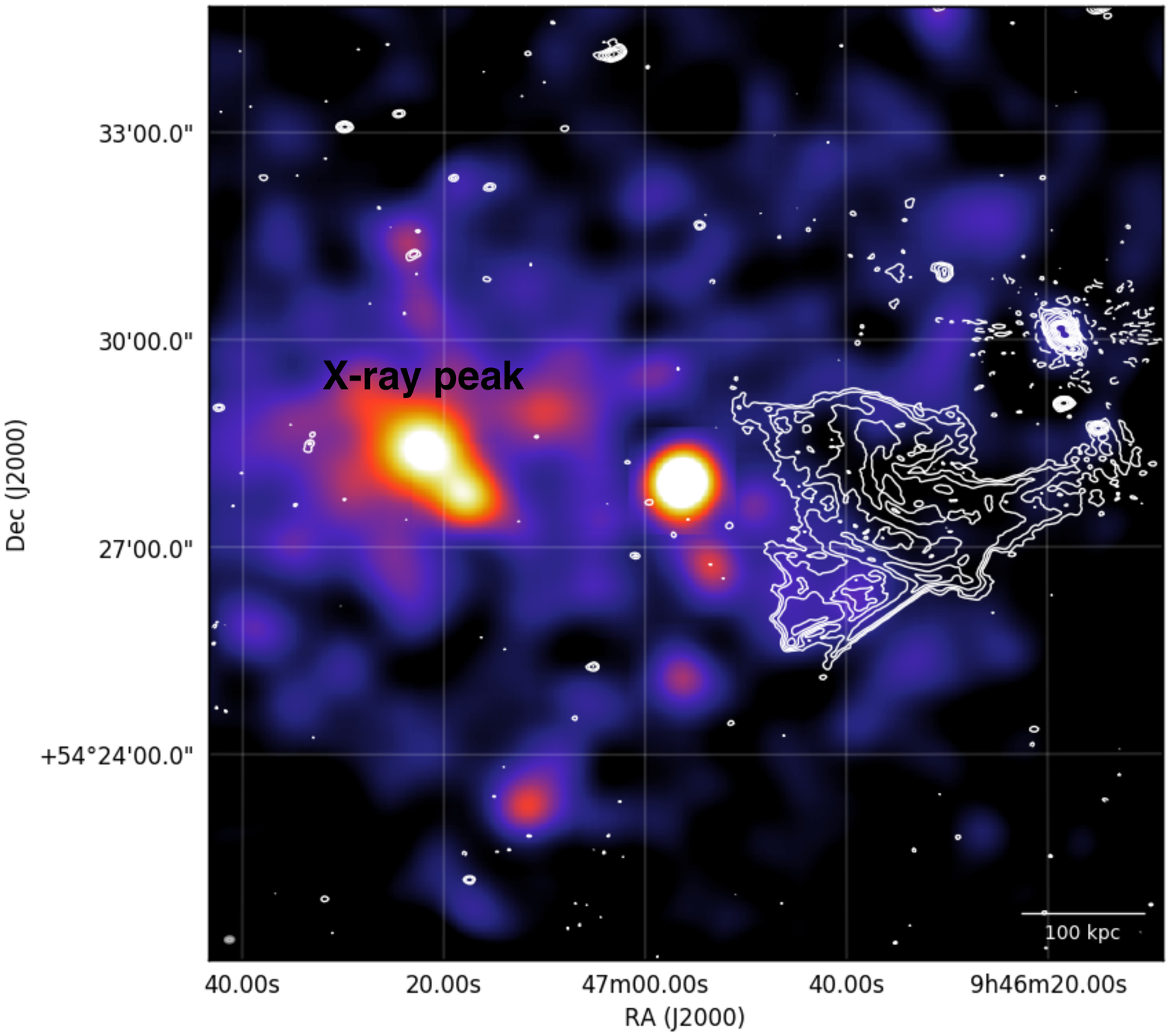}
}
\resizebox{0.49\hsize}{!}{
\includegraphics[angle=0]{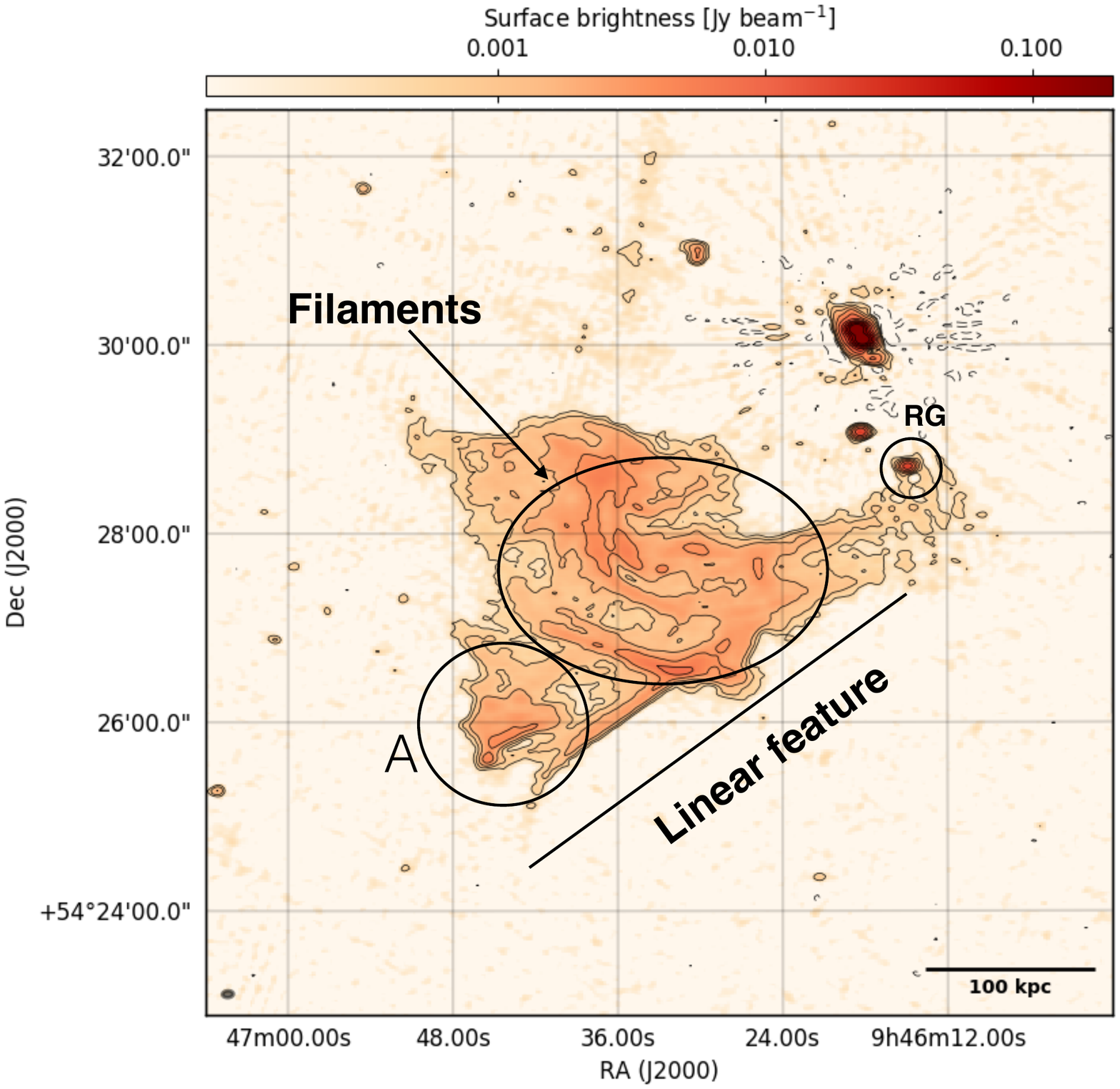}
}
\resizebox{0.49\hsize}{!}{
\includegraphics[angle=0]{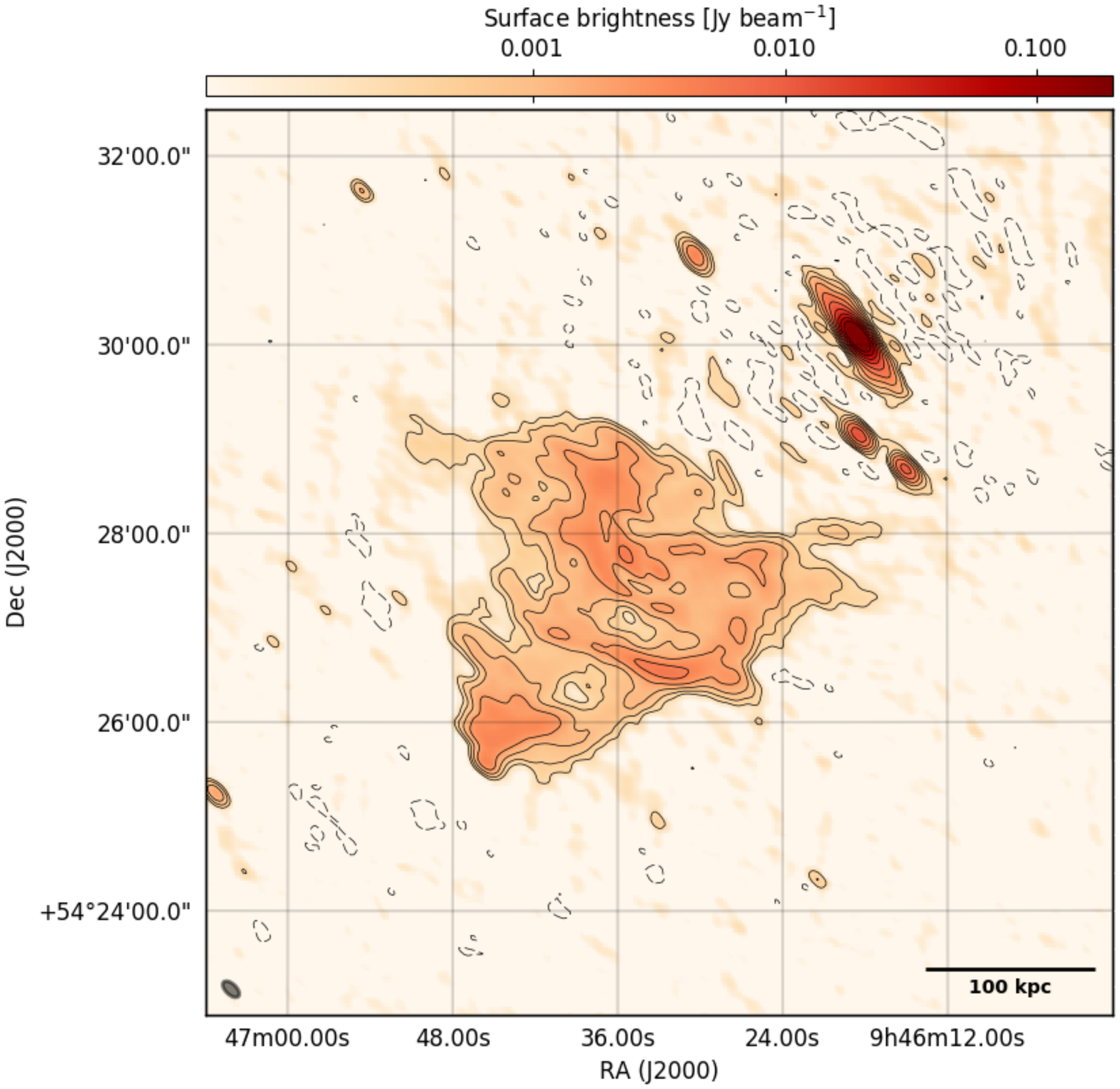}
}
\resizebox{0.49\hsize}{!}{
\includegraphics[angle=0]{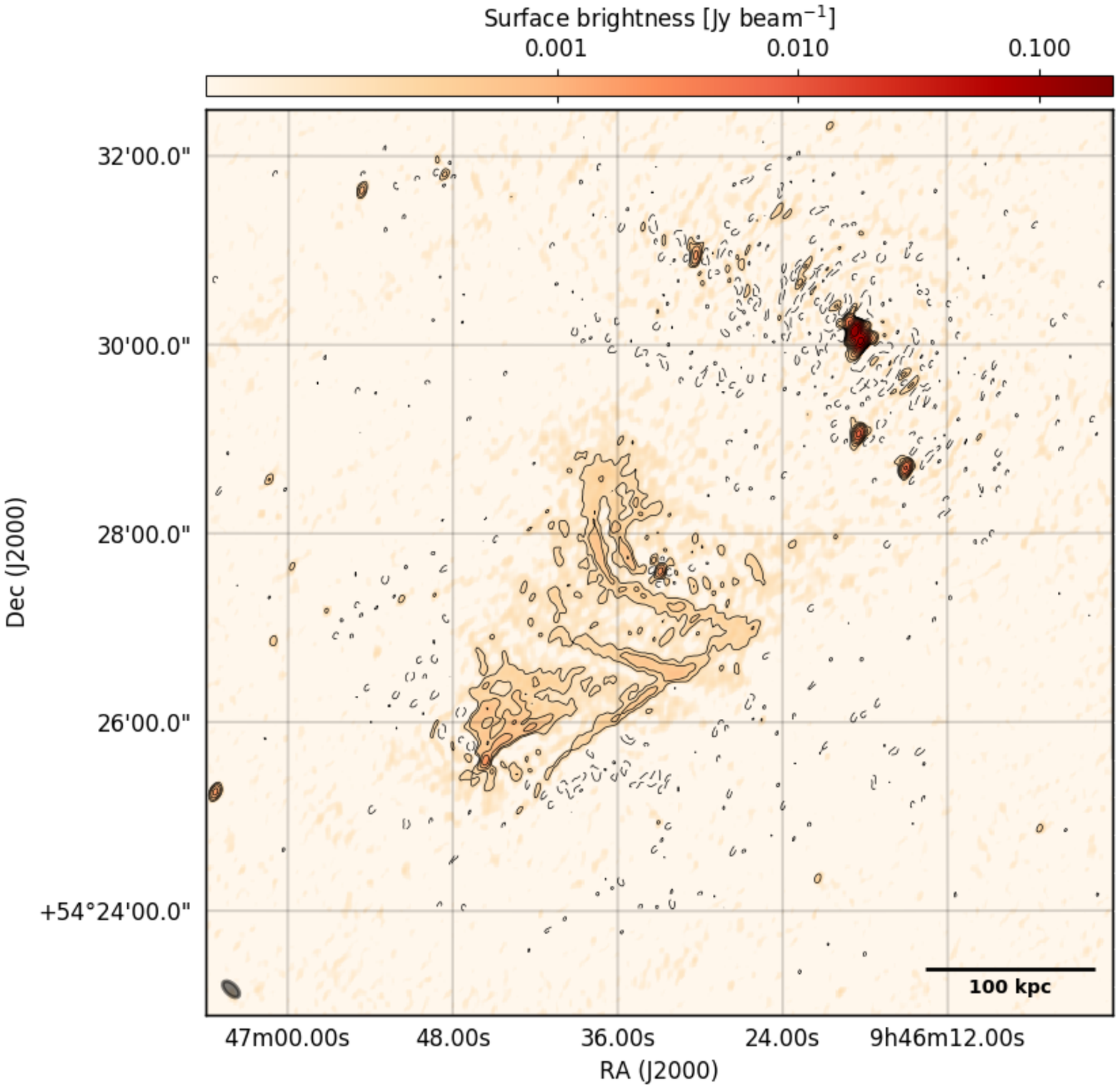}
}
\resizebox{0.49\hsize}{!}{
\includegraphics[angle=0]{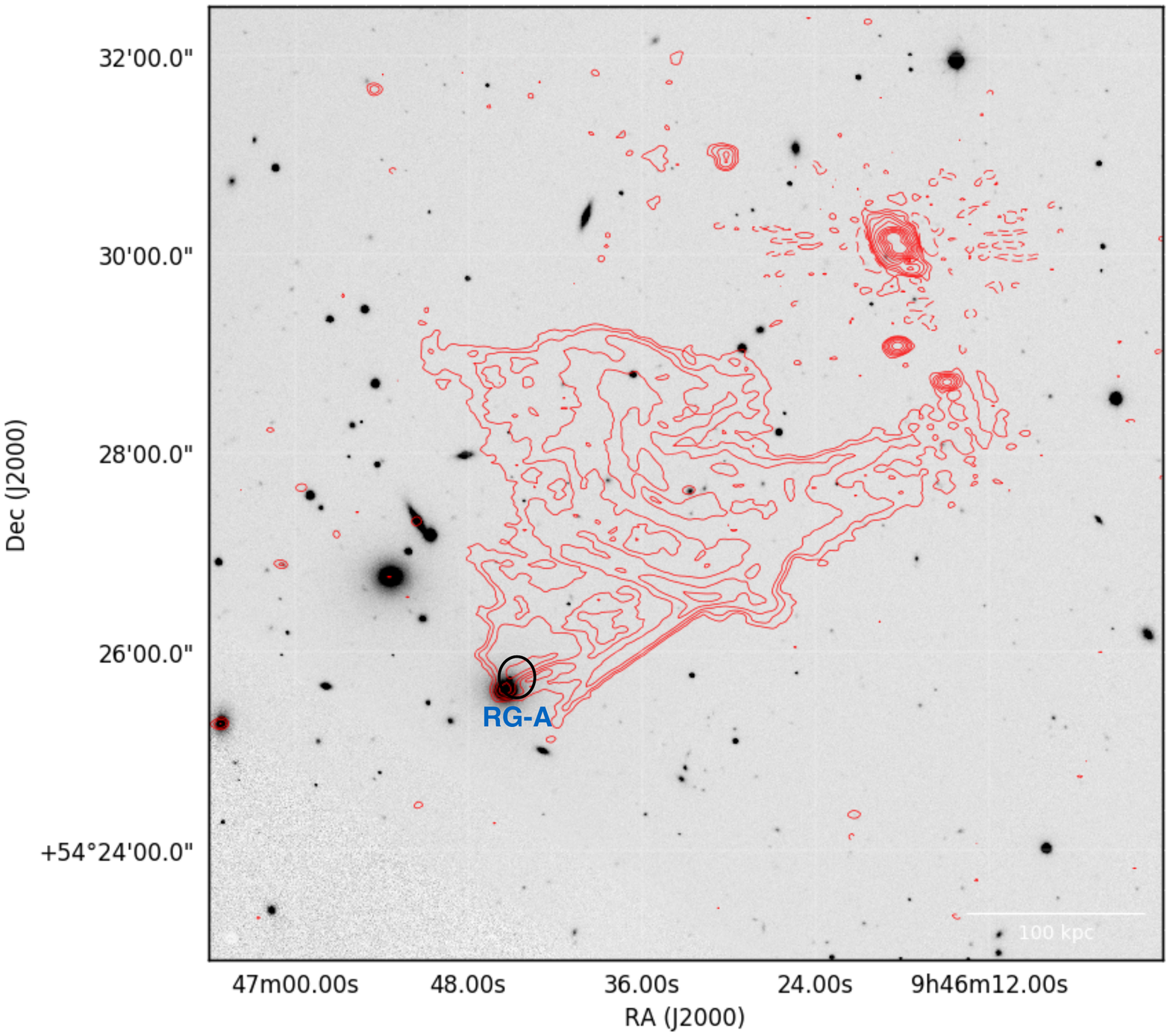}
}
\resizebox{0.49\hsize}{!}{
\includegraphics[angle=0]{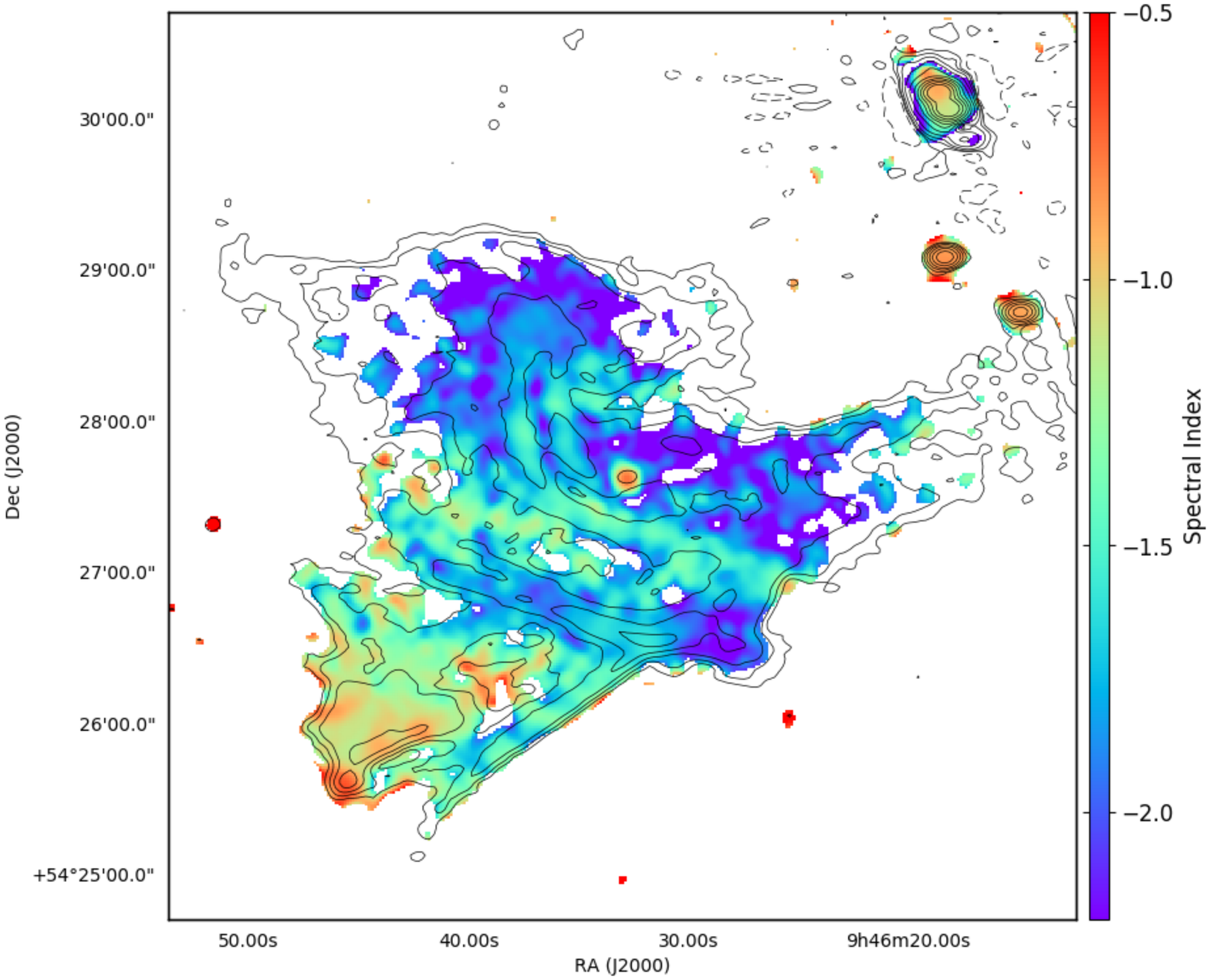}
}
\caption{
\textit{Top left} panel: Full-resolution LOFAR image contours (white; as shown in the \textit{right} panel) of SDSS-C4-DR3-3088, overlaid on an exposure-corrected \textit{Chandra} image in the 0.5-2.0 keV energy band with a total integration time of 17ks.
\textit{Top right} panel: The 150 MHz LOFAR image of SDSS-C4-DR3-3088 (\asec{6} $\times$ \asec{6}) where the black contours and dashed lines show the (1,2,4,...)$\times$5$\times$ $\sigma_{\rm{LOFAR150}}$ and -3$\times$ $\sigma_{\rm{LOFAR150}}$ levels respectively, where $\sigma_{\rm{LOFAR150}}$=80$\mu$Jy/beam.  
\textit{Middle left} panel: The 325 MHz GMRT image (\asec{13} $\times$ \asec{7}) where the black contours and dashed lines show the (1,2,4,...)$\times$5$\times$ $\sigma_{\rm{GMRT325}}$ and -3$\times$ $\sigma_{\rm{GMRT325}}$ levels respectively, where $\sigma_{\rm{GMRT325}}$=43$\mu$Jy/beam. 
\textit{Middle right} panel: The 610 MHz GMRT image (\asec{6} $\times$ \asec{4}) where the black contours and dashed lines show the (1,2,4,...)$\times$5 $\times$ $\sigma_{\rm{GMRT610}}$ and -3$\times$ $\sigma_{\rm{GMRT610}}$ levels respectively, where $\sigma_{\rm{GMRT610}}$=40$\mu$Jy/beam. \textit{Bottom left} panel: The LOFAR 150~MHz contours as shown in left panel in \textit{red} overlaid on an SDSS r-band image of SDSS-C4-DR3-3088. {\color{black}{Possible optical counterpart is marked as RG-A.}} 
\textit{Bottom right} panel: The LOFAR 150~MHz contours as shown in the top \textit{right} panel overplotted on the high resolution (\asec{7} $\times$ \asec{7}) spectral index map of SDSS-C4-DR3-3088. 
}
\label{fig:sdssc4}
\end{center}
\end{figure*} 

%----about the cluster-----% 
The galaxy cluster SDSS-C4-DR3-3088 is located at a redshift of $z = 0.046$. 61 cluster members were spectroscopically confirmed by \cite{simard09}. The top-left panel of Figure~\ref{fig:sdssc4} shows the 150~MHz LOFAR image contours overlaid on our \textit{Chandra} X-ray image. {\color{black}{The peak in the diffuse X-ray emission resides towards the east of the radio source at a (large) distance of 500~kpc (marked as the `X-ray peak'). We note that this peak is due to the contribution from a point source (CGCG 265-040, $z=0.047$; \citealt{yang07}). Since the X-ray emission is irregular in morphology, it could suggest that the cluster could be dynamically disturbed.}}

%----radio morphology-----%
The top-right panel shows the 150~MHz LOFAR image of the radio source. The middle-left and -right panels show the GMRT 325 and 610~MHz images, respectively. In all images, the radio source has an irregular, filamentary morphology {\color{black}{with a resemblance of a wide-angle radio tail.}}
Regions of interest have been marked on the LOFAR 150~MHz image. Starting from region~A, the source extends about 400~kpc in NW direction. There is a linear feature along the edge of the source 500~kpc in length. In the LOFAR 150~MHz image, this linear feature appears to be connected to a compact source (marked as RG). However, this extension of the diffuse emission is not recovered in the GMRT 325~MHz and 610~MHz images, suggesting that it is steep-spectrum emission. The central `Filaments' region is detected at all radio frequencies. {\color{black}{The radio source gives of an oddly shaped radio tail wedge with RG-A in its focus. The morphological connection of RG-A with the larger-scale radio emission is persistent across all radio frequency bands, which strongly suggests a physical connection.}}
%----optical information-----%
The bottom-left panel of Figure~\ref{fig:sdssc4} shows the SDSS r-band image of the cluster overlaid with the LOFAR 150~MHz contours in red. We do not identify any optical counterpart to the compact radio source RG (see top-right panel). Given the fairly low redshift of the cluster (z = 0.046), this likely means that the compact radio source is a background object at higher redshift, and the apparent connection with the diffuse emission at 150~MHz is a chance alignment. We find an optical galaxy at the location of RG-A, which we identify as its host. The brightest cluster galaxy is located near the peak in the X-ray emission (outside the extend of the optical image). 

%----spectral index-----%
To measure the integrated spectral index of the source, a set of new images were made with the same \textit{uv}-range in order to match same spatial scales in all available radio frequencies (see Section \ref{sec:spix_maps} for details). The measured integrated flux densities of the source over exactly the same area (defined by the 3$\sigma$ contours of the LOFAR image) at 150~MHz, 325~MHz and 610~MHz are $S_{150}$ = 1.83 $\pm$ 0.18 Jy, $S_{325}$= 0.45 $\pm$ 0.50 Jy, $S_{610}$=0.16 $\pm$ 0.02 Jy. A single power-law fit gives an integrated spectral index measurement of $\alpha_{\rm{150-610}}$ = -1.74 $\pm$ 0.23. We have also used the NVSS image to calculate the total flux density and adding this extra frequency point, the single power-law fit gives an integrated spectral index measurement of $\alpha_{\rm{150-1400}}$ = -2.00 $\pm$ 0.06. This suggests, there could be a hint of curvature at the higher frequency. To create a high-resolution (\asec{7}) spectral index map (same method as Section 3.4), we only used the LOFAR 150~MHz and 610~MHz (excluding the GMRT 325~MHz map) which is shown in Figure~\ref{fig:sdssc4} bottom right panel. The spectral index at the location of the optical host galaxy RG-A is -0.8. The spectral index steepens in NW direction over a distance of 100~kpc (region~A) to a value of -1.3. The filaments have an overall steep spectral index distribution flatter than the surrounding diffuse emission. The Linear feature has a spectral index value ranging from -1.3 to -1.5 prior to the `Filaments' region. The diffuse emission apparently connecting to the source RG (Figure~\ref{fig:sdssc4} \textit{left} panel) has the steepest spectral index with a value of -3.0 {\color{black}{(See Appendix B for additional spectral index maps).}}

\subsection{Abell~2048} 
\label{section:results_abell2048}

\begin{figure*}[!htb]
\begin{center}
\resizebox{0.49\hsize}{!}{
\includegraphics[angle=0]{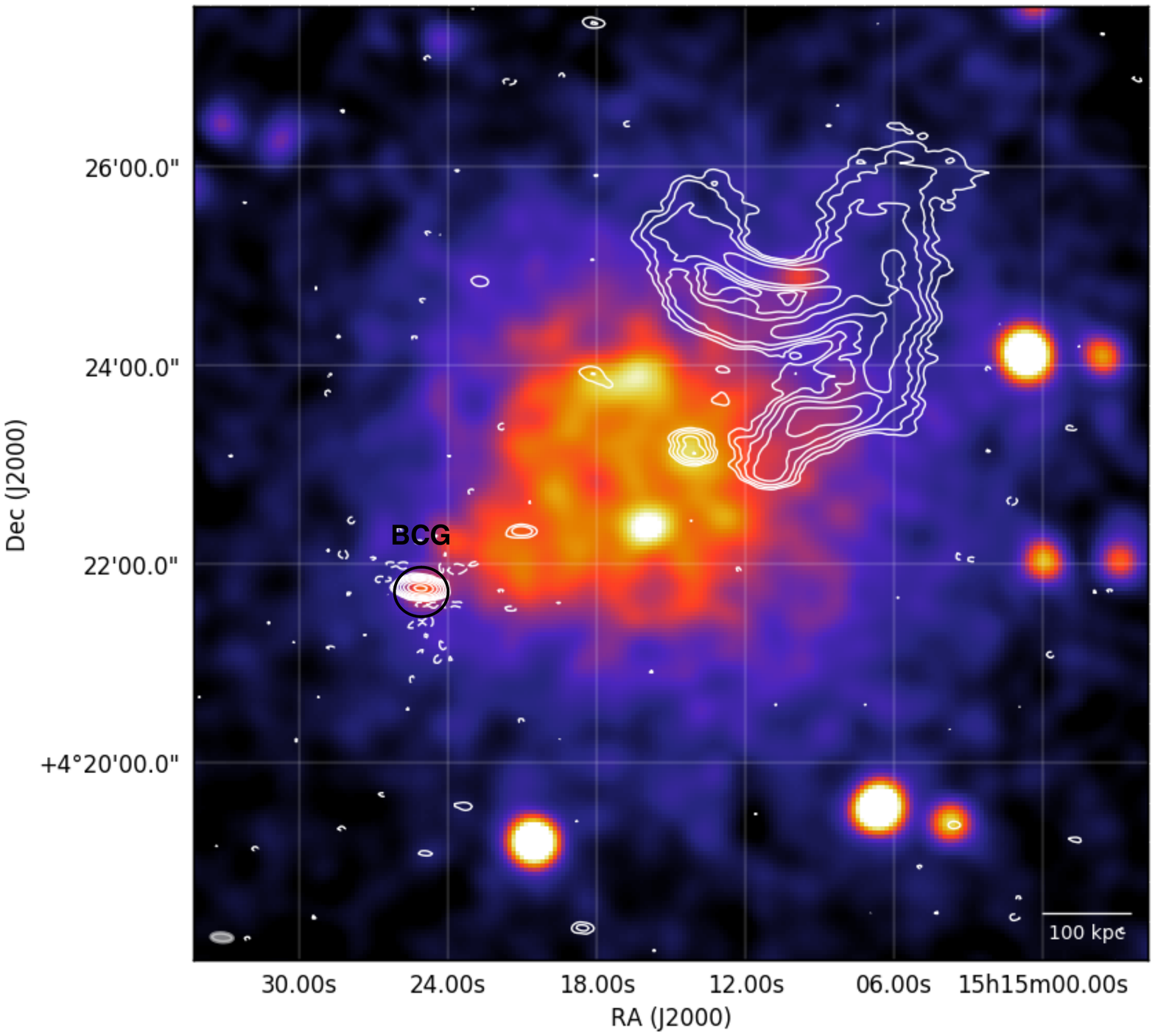}
}
\resizebox{0.49\hsize}{!}{
\includegraphics[angle=0]{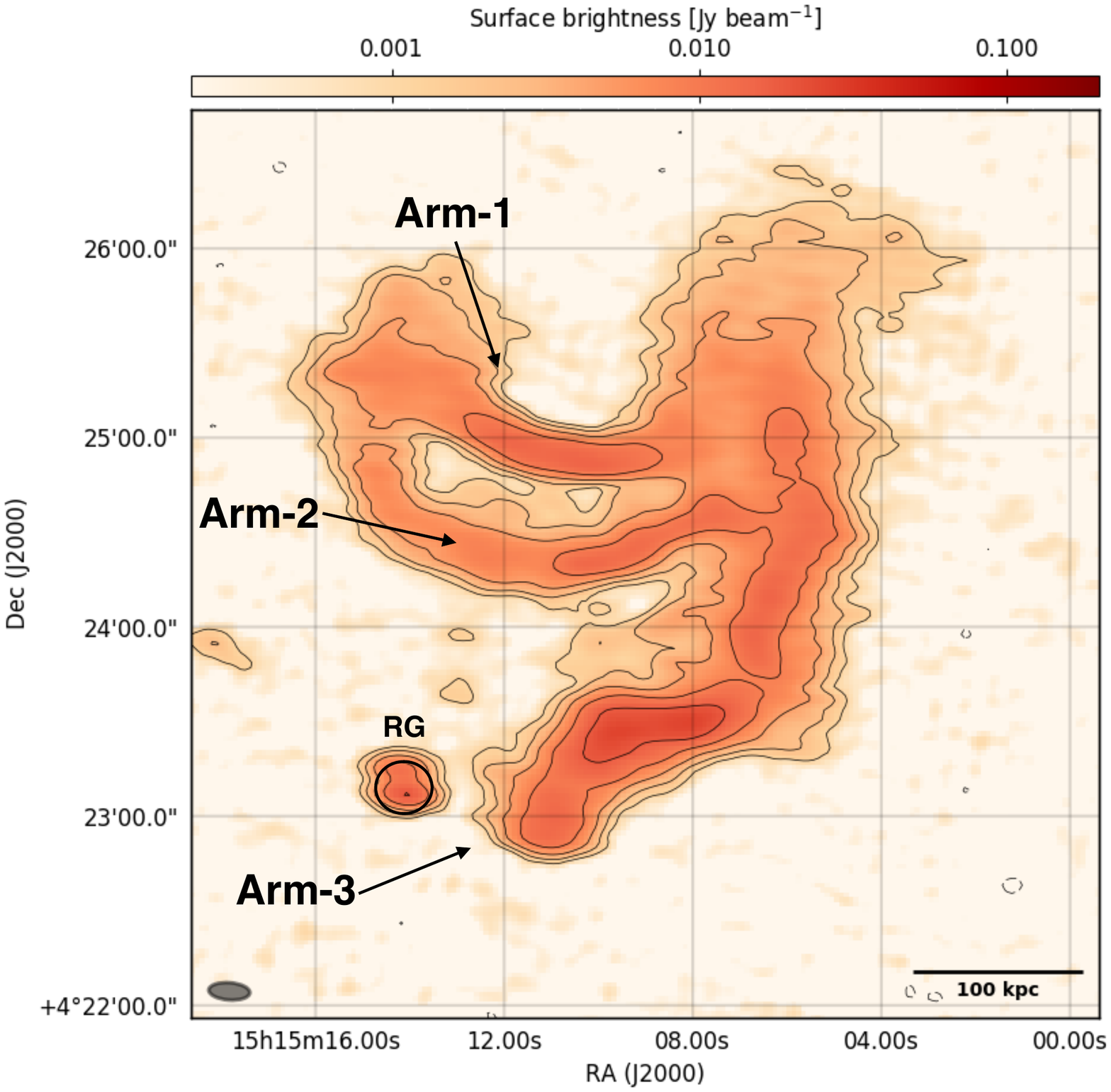}
}
\resizebox{0.49\hsize}{!}{
\includegraphics[angle=0]{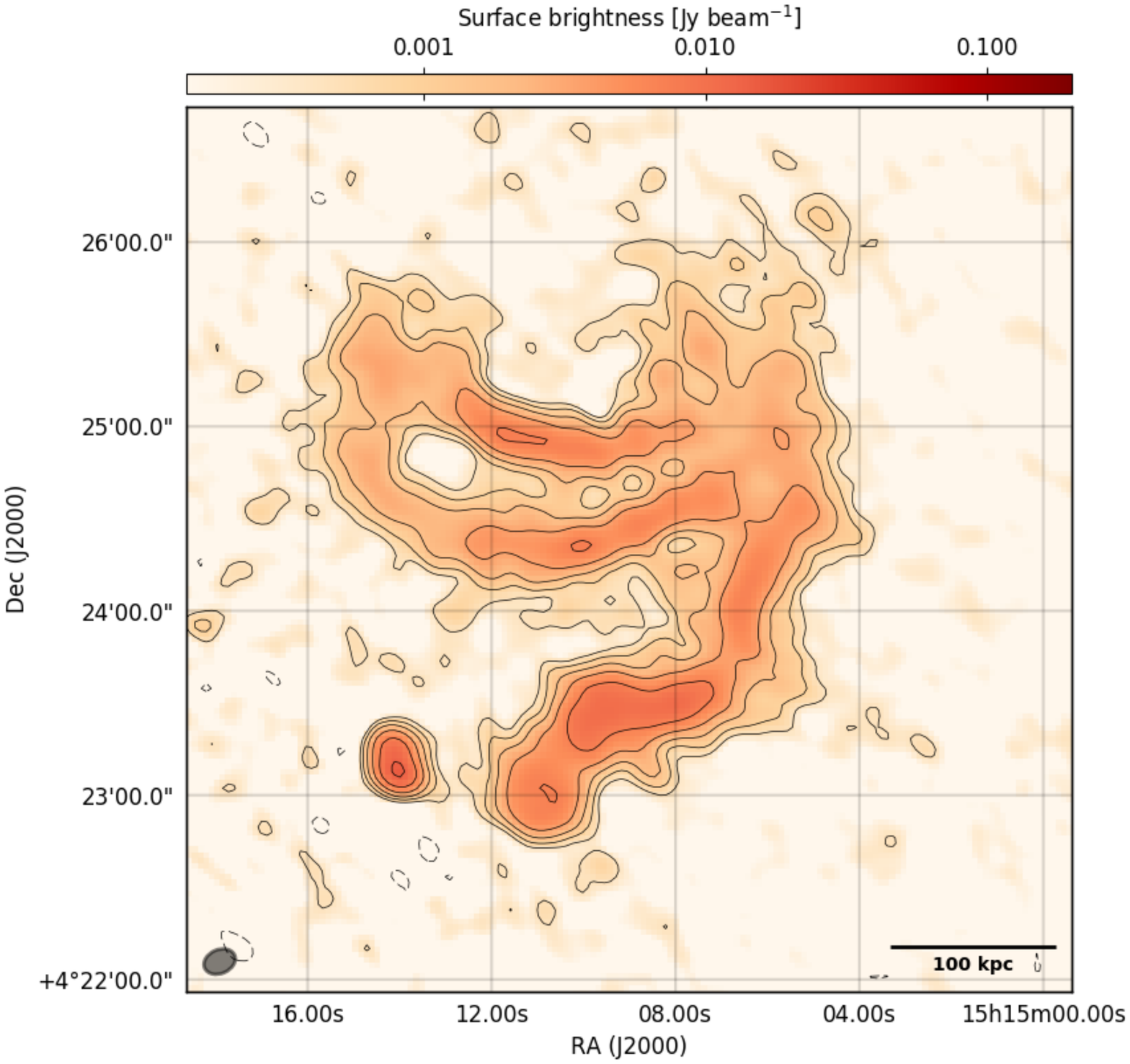}
}
\resizebox{0.49\hsize}{!}{
\includegraphics[angle=0]{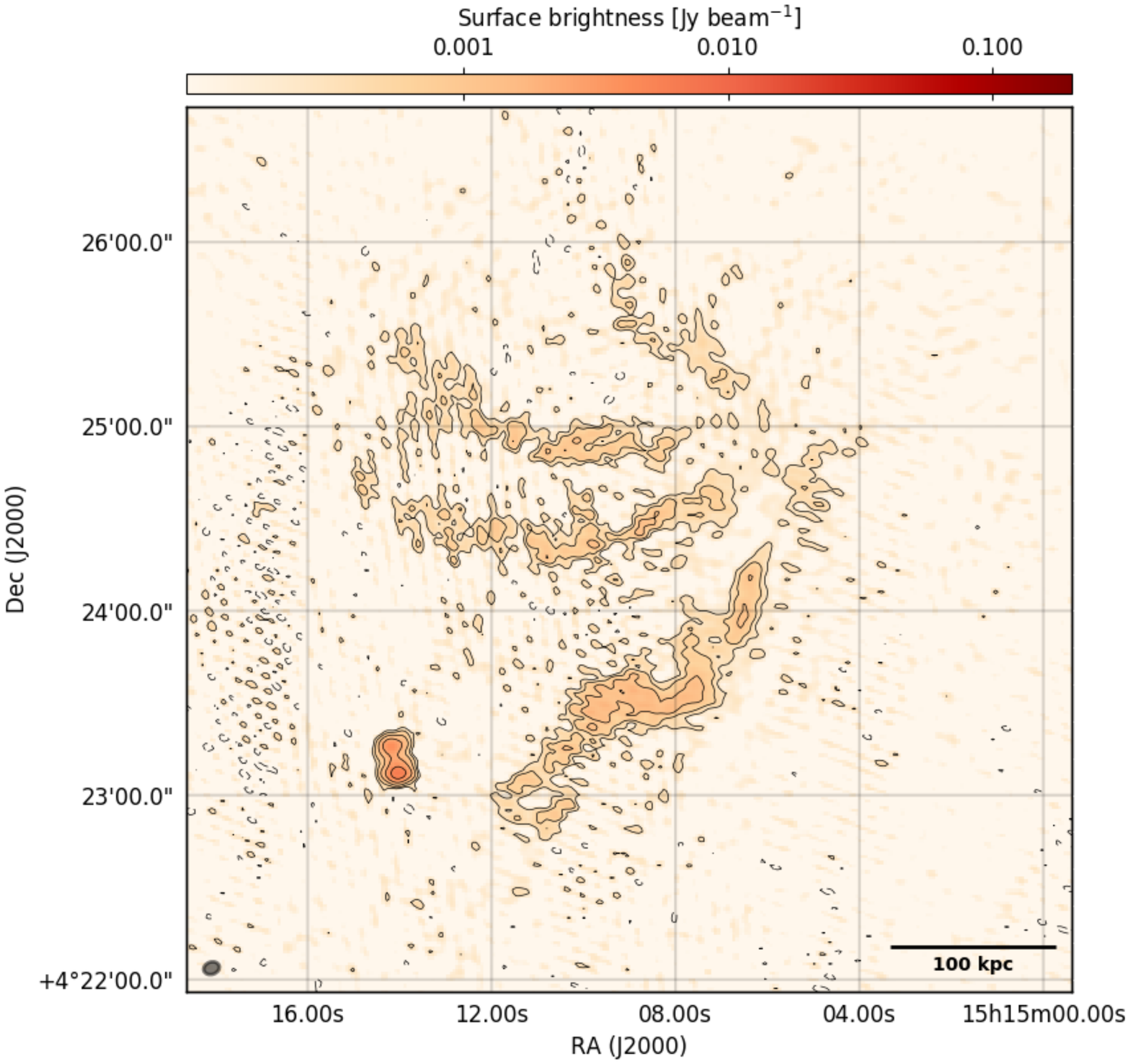}
}
\resizebox{0.49\hsize}{!}{
\includegraphics[angle=0]{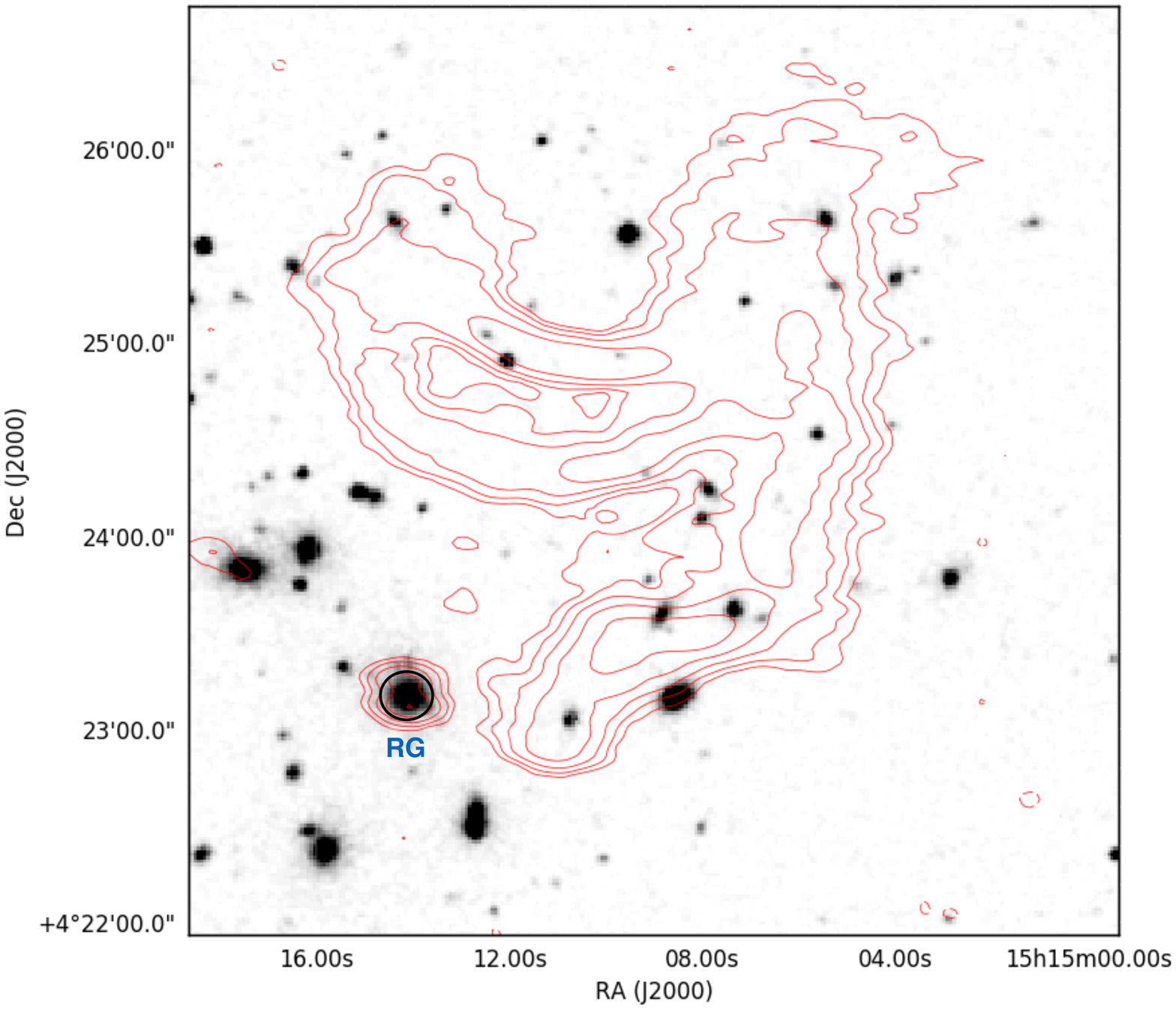}
}
\resizebox{0.49\hsize}{!}{
\includegraphics[angle=0]{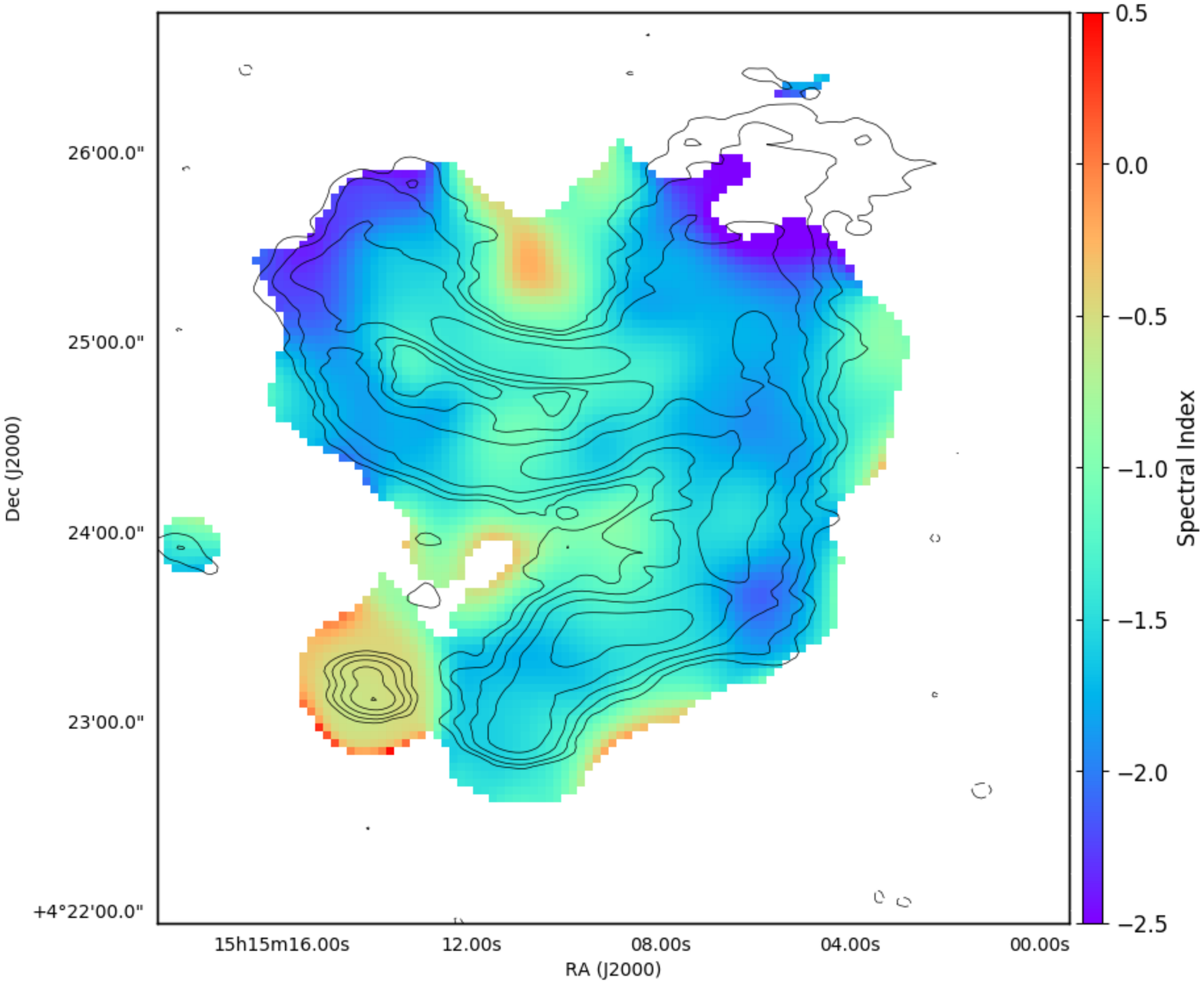}
}
\caption{
\textit{Top left} panel: Shows the full-resolution LOFAR image contours (white; as shown in the top \textit{right} panel) of Abell~2048, overlaid on an exposure-corrected, background-subtracted \textit{XMM-Newton} image in the 0.5-2.0 keV energy band with a total integration time of 68 ks. 
\textit{Top right} panel: The 150 MHz LOFAR image (\asec{13} $\times$ \asec{6}) of the source Abell~2048; where the black contours and dashed lines show the (1,2,4,...)$\times$5$\times$ $\sigma_{\rm{LOFAR150}}$ and -3$\times$ $\sigma_{\rm{LOFAR150}}$ levels respectively, where $\sigma_{\rm{LOFAR150}}$=266$\mu$Jy/beam.
\textit{Middle left} panel: The 325 MHz GMRT image  (\asec{11} $\times$ \asec{8}) of Abell~2048; where the black contours and dashed lines show the (1,2,4,...)$\times$5 $\times$ $\sigma_{\rm{GMRT325}}$ and -3$\times$ $\sigma_{\rm{GMRT325}}$ levels respectively, where $\sigma_{\rm{GMRT325}}$=125$\mu$Jy/beam. 
\textit{Middle right} panel: The 610 MHz GMRT image (\asec{8} $\times$ \asec{5}) of Abell~2048 where the black contours and dashed lines show the (1,2,4,...)$\times$5 $\times$ $\sigma_{\rm{GMRT610}}$ and -3$\times$ $\sigma_{\rm{GMRT610}}$ levels respectively, where $\sigma_{\rm{GMRT610}}$=117$\mu$Jy/beam. 
\textit{Bottom left} panel: The LOFAR 150~MHz contours as shown in the \textit{left} panel in \textit{red} overlaid on an DSS2 r-band image of Abell~2048. A possible optical counterpart is marked as RG.  
\textit{Bottom right} panel: The LOFAR 150~MHz contours as shown in the top \textit{right} panel overplotted on the high resolution (\asec{20} $\times$ \asec{20}) spectral index map of Abell~2048.}
\label{fig:a2048}
\end{center}
\end{figure*} 

%----about the cluster----%
Abell~2048 is a galaxy cluster located at a redshift of $z = 0.0972$ \citep{strublerodd99}. The top-left panel of Figure~\ref{fig:a2048} shows the LOFAR 150~MHz contours (in white) overplotted on a background-corrected \textit{XMM-Newton} X-ray image. The brightest cluster galaxy of the system is marked as BCG (and lies outside the extend of the radio and optical images). {\color{black}{The morphology of the X-ray emission is slightly elongated along the NW-SE direction and is not strongly peaked at the centre. This could indicate that the dynamical state of this cluster is disturbed.}} 

%----radio morphology----%
The steep-spectrum radio source is located at the outskirts of the cluster. The top-right panel in Figure~\ref{fig:a2048} shows the 150~MHz LOFAR image of the filamentary radio source with a complex morphology. The largest extent of the source is 400~kpc. Three distinctive `arm'-like features are marked in the same Figure. {\color{black}{There is a hint of a connection between the extension of `Arm-3' and the compact source which is  marked as `RG'}}. In the middle panel, we show the GMRT 325~MHz and GMRT 610~MHz images on the left and right panel, respectively. At 610~MHz, most of the diffuse emission in the `arms' is faint, suggesting it has a very steep radio spectrum. \cite{vanweeren11} presented the VLA 1.4~GHz image in `C' configuration which confirms the steep nature of the diffuse emission, and set an upper limit on the polarization fraction of 8\% for the source. 

%----optical----%
The bottom-left panel shows the optical DSS2 r-band image of the cluster overlaid with the LOFAR 150~MHz contours in red. We have identified the elliptical galaxy MCG+01--39--011 (z = 0.095) as the host of the radio source `RG' based on accurate co-location. There are no other obvious optical counterparts associated with the radio emission in the arms, although several cluster galaxies lie within the arm areas.

%----spectral index----%
In order to measure the spectral properties of the radio source in Abell~2048, we used the same \textit{uv}-range to re-image the dataset in all available radio frequencies (see Section \ref{sec:spix_maps} for details). The measured integrated flux densities of the source are $S_{150} = 1.57 \pm 0.25$~Jy, $S_{325} = 0.50 \pm 0.05$~Jy, $S_{610} = 0.17 \pm 0.02$~Jy as measured at 150~MHz, 325~MHz and 610~MHz, respectively. A single power-law fit gives an integrated spectral index measurement of $\alpha_{\rm{150-610}} = -1.59 \pm 0.17$. We also re-calculated the integrated spectral index value by adding the 1.4~GHz VLA total flux density measurement (from \citealt{vanweeren11}). The resulting value of $\alpha_{\rm{150-1400}} = -1.94 \pm 0.20$ suggests that the radio source possibly has a curved spectrum. We re-imaged the LOFAR and GMRT observations to make a high-resolution spectral index map, which is shown in Figure~\ref{fig:a2048} (bottom left panel). The value of the spectral index is uniformly steep (on average) in the `arm' regions with $\alpha = -1.65 \pm 0.10$ with small variations. No particular trend (flattening or steepening) of the spectral index is noticed along any direction. The spectral index of the source RG is flat with a value of $\alpha = -0.8$. 
%===============================================================================

\section{Discussion}
\label{section:discussion}

In this paper, we have presented a multi-wavelength study of three ultra-steep spectrum diffuse radio sources in galaxy clusters. {\color{black}{The morphology of these sources do not fit with a scenario in which they are mere aged confined bubbles from AGNs in the ICM. Also, the spectral index shows patchy structures with no clear trend within the source. In addition, the kinematic age (travel time of the electrons in the ICM) of the source ($\sim$Gyr) is much larger than the radiative life-time ($\sim$Myr) of the electrons; which suggests re-energisation of the electrons occurs at the location. Although limited in numbers, this study aims to explore properties of this relatively unexplored class of sources, to look at possible common characteristics and the connection with the dynamical properties of the hosting cluster. Below, we discuss the common physical properties of these sources based from these observations.}}

\subsection{AGN connection}

All these sources show a very filamentary, irregular morphology at low radio frequencies.  As shown in the results section, these sources are morphologically connected with an AGN (e.g.: RG1-A, RG2-A for Abell~2593; RG-A for SDSS-C4-DR3-3088 and RG for Abell~2048). Spectral index values at these locations are on the order of -0.7 which is in agreement with the injection spectral index of an AGN, and the spectral index tends to steepen along the tail of the radio galaxy.

\subsection{\color{black}{Spectral index}}

The three sources have an overall ultra-steep radio spectrum. The relatively strong low-frequency emission traces old population of electrons, as high-energy electrons age due to synchrotron and IC losses and are visible only for tens of Myrs. Old lobes of radio galaxies (AGNs) are one of the main candidates for these fossil plasma. The measured integrated spectral index between 150~MHz, 325~MHz and 610~MHz is less steep than the one measured in between 150~MHz, 325~MHz, 610~MHz and 1400~MHz {\color{black}{(as stated in Section 4.1)}}, which possibly suggests that there is curvature towards higher frequencies. Deep observations at 1.4~GHz are needed in order to confirm this. It is important to note that, all the sources have non-uniform spectral index across the source. This could suggest that there is a possible mix of cosmic-ray populations with different ages, losses and re-acceleration efficiencies. Therefore, the re-energisation of these particles can also be different across the source. So, even though there is a hint of curvature in the integrated spectral index, the sources should not be interpreted as a reservoir of homogeneous cosmic-ray population. {\color{black}{Also, we note that the integrated spectral indices of these sources are much steeper than the classical radio relics \citep{feretti12}.}}   

\subsection{Mass, temperature, dynamical state and position of the radio source in the cluster}

In Table~\ref{tab:xray-parameter} we have listed the derived mass and the global temperature of each galaxy cluster, which shows that these are only moderately massive ($M_{500} <$ $10^{15}$ $M_\odot$).
{\color{black}{Most of the previously discovered radio phoenix candidates (apart from the one in Abell~1914; \citealt{mandal18}) tend to be located in less massive system of galaxy clusters.}}
From the same table, we see that the radio sources reside well within the $R_{500}$ value and extend about 500~kpc. This means that the radio sources are located within the ICM towards the cluster centre. Deeper X-ray observations are needed in order to map the temperature in different regions and characterise the merger dynamics more accurately. {\color{black}{However, the apparent disturbed morphologies of these clusters suggest that the systems could be unrelaxed and undergoing minor mergers.}} Although, radio phoenix candidates with highly disturbed clusters also do exist (e.g., Abell 1914 and Abell 2443; \citealt{mandal18}, \citealt{clarke13}) but a clear connection of a shock with radio phoenix is still missing.

\subsection*{\color{black}{Interpretation}} 

Ultra-steep spectrum radio sources in clusters with complex, filamentary morphologies have sporadically been seen in previous studies (e.g.: \citealt{slee01}, \citealt{kempner04}) and were characterised as \textit{relic or ghost radio galaxies} and proposed to name these radio phoenices. The origin of these sources was speculated to be AGNs. \cite{slee01} showed that the estimated travel time of the brightest cluster galaxies are much longer than the modelled ages of relic radio sources, and a nearby bright elliptical galaxy always provided a decent match as a candidate source of origin. Based on our observations and results, the sources presented in this paper also appear to belong to the same class of objects.

Shocks and ICM motion can affect the morphology of ghost plasma/bubbles. Recent simulations of cluster radio galaxy tails, passing through ICM shocks or with large relative motions, show filamentary morphologies, ultra-steep and curved spectra (\citealt{nolting19}) similar to what is observed. Earlier it was already proposed that shocks can compress the radio plasma if this is still poorly mixed with the ICM. This compression re-energizes the electrons to boost their visibility at frequencies below few hundred MHz (\citealt{eg01}, \citealt{ensslinbruggen02}).

Another scenario for the formation of these ultra-steep spectrum sources in clusters invokes shock re-acceleration of fossil plasma through the Fermi-I mechanism (DSA; e.g: \citealt{kangryu15}). This will flatten the curved radio spectrum and if the shock is strong enough, the spectrum follows a power-law distribution, as is typically the case for giant radio relics. Contrary to re-acceleration, compression only shifts the spectrum in the Flux-Frequency space without changing the underlying shape.

For our sources we found spatially non-uniform spectral indices. Together with the observed filamentary morphologies suggest a different degree of mixing of the relativistic particles from the AGNs with the ICM. This means that the compression scenario proposed by \cite{eg01} does not explain all the aspects. The inhomogeneous mixing of relativistic particles with the ICM implies an averaging effective adiabatic index ((4/3)<$\gamma_{ad}$<(5/3)) value which consequently lead to different effective compression ratios across the source. Under these conditions and driven by a complex cluster dynamics, plasma instabilities may also play an important role in the re-acceleration of particles within the radio sources. These mechanisms cannot be modelled by existing numerical simulations. However, recently, with the help of low-frequency (sub-GHz) observations,  \citealt{gasperin17} showed the possibility of an interplay between fossil plasma and the perturbed ICM which can gently re-accelerate relativistic particles injected by AGNs of the galaxy cluster Abell~1033. If this kind of `gentle-reacceleration' mechanism is common for aged plasma in clusters of galaxies, it could provide seed population of relativistic electrons and possible connection between radio galaxies and re-acceleration mechanisms in radio halos or radio relics. An example of this scenario could be the Abell 3411-3412 galaxy cluster pair, where \citealt{vanweeren17} showed a morphological connection between a radio galaxy and a radio relic.

%===============================================================================

\section{Conclusion}
\label{section:conclusions}

In this paper, we studied three ultra-steep spectrum sources in galaxy clusters based on radio and X-ray observations. We found that these three sources i) have complex filamentary radio morphology, ii) show hints of curved spectra, and iii) are probably related to AGN fossil radio plasma. Based on these properties, we conclude that these sources do belong to the category of `radio phoenices', as was defined by \cite{kempner04}. 

In this work we found that these three sources are located in galaxy clusters with low-mass (<$10^{15} M_\odot$) and with an un-relaxed dynamical state. This provides support for the scenario that phoenices are connected with shocks and/or ICM motions. In addition, we determined that these sources have spatially non-uniform spectral indices. This suggests a different degree of mixing of the relativistic particles from the AGNs with the ICM and implies that several mechanisms are operating for the re-energisation of the plasma.

With the advancement of low-frequency radio telescopes and data processing techniques, upcoming high-resolution and sensitive surveys at or below 150~MHz (such as LoTSS at 150~MHz; \citealt{shimwell19}, \asec{6}, ~100$\mu$Jy/beam; LoLSS at 50~MHz; de Gasperin et al. in prep, \asec{20}, 1.5 mJy/beam) will be excellent tools to identify many more revived fossil plasma sources in galaxy clusters, enabling statistical analysis of larger samples.

%===============================================================================

\section*{Acknowledgments}

We thank the anonymous referee for useful comments. This paper is based (in part) on data obtained with the International LOFAR Telescope (ILT) under project code LC9\_027 and LC6\_015. LOFAR (\citealt{vanhaarlem13}) is the Low Frequency Array designed and constructed by ASTRON. It has observing, data processing, and data storage facilities in several countries, that are owned by various parties (each with their own funding sources), and that are collectively operated by the ILT foundation under a joint scientific policy. The ILT resources have benefitted from the following recent major funding sources: CNRS-INSU, Observatoire de Paris and Université d'Orléans, France; BMBF, MIWF-NRW, MPG, Germany; Science Foundation Ireland (SFI), Department of Business, Enterprise and Innovation (DBEI), Ireland; NWO, The Netherlands; The Science and Technology Facilities Council, UK ; Istituto Nazionale di Astrofisica (INAF), Italy. We thank the staff of the GMRT that made these observations possible. GMRT is run by the National Centre for Radio Astrophysics of the Tata Institute of Fundamental Research. This paper is based on the data obtained with the International LOFAR Telescope (ILT). The Leiden LOFAR team acknowledge support from the ERC Advanced Investigator programme NewClusters 321271 and the VIDI research programme with project number 639.042.729, which is financed by the Netherlands Organisation for Scientific Research (NWO). MJH acknowledges support from STFC [ST/R000905/1]. This research made use of APLpy, an open-source plotting package for Python hosted at \href{url}{http://aplpy.github.com.} 

%===============================================================================

\bibliographystyle{aa} 
\bibliography{paper}

\clearpage
\onecolumn
\begin{appendix}

\section{Spectral index error map}

\begin{figure*}[!htb]
\begin{center}
\resizebox{0.49\hsize}{!}{
\includegraphics[angle=0]{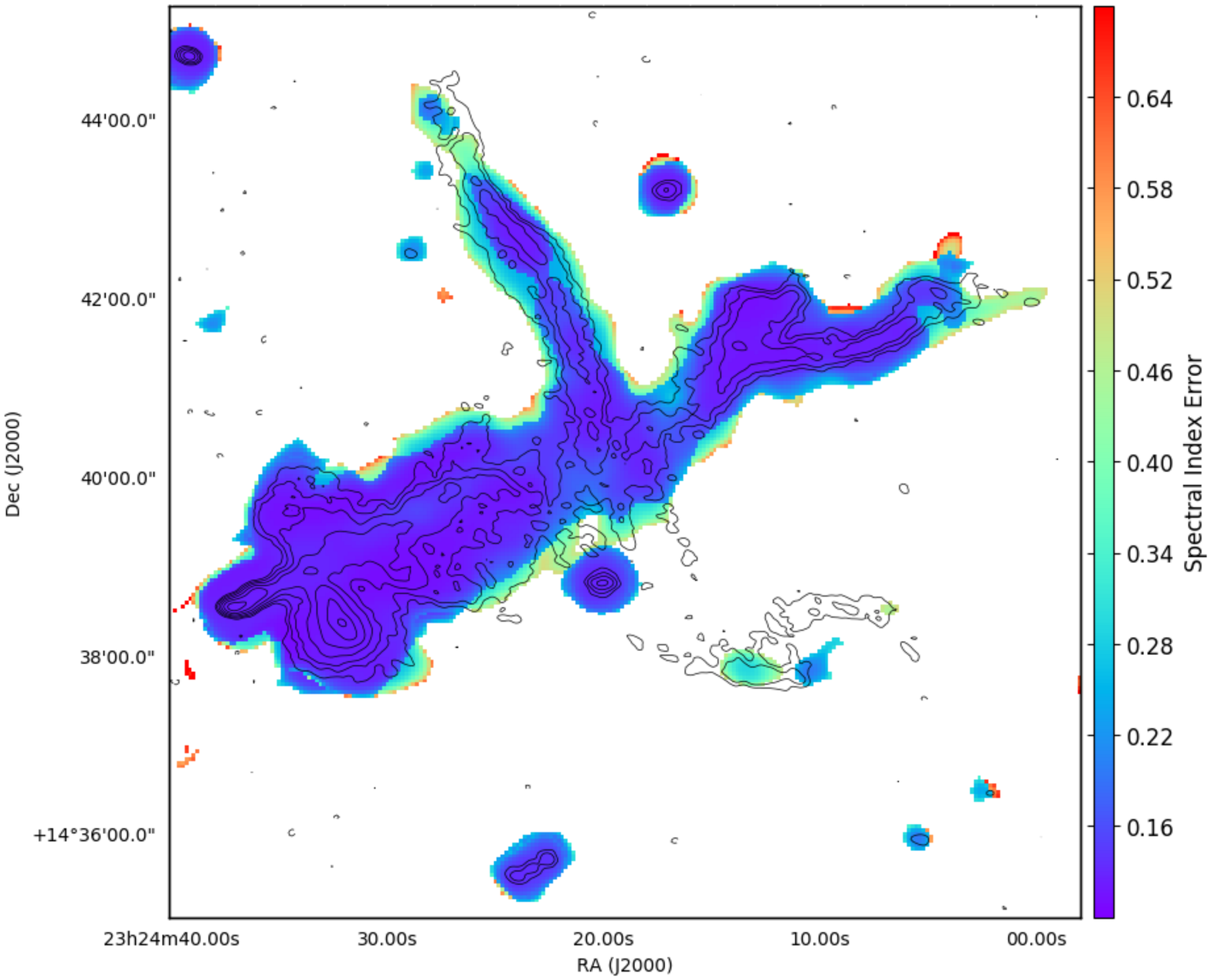}
}
\resizebox{0.49\hsize}{!}{
\includegraphics[angle=0]{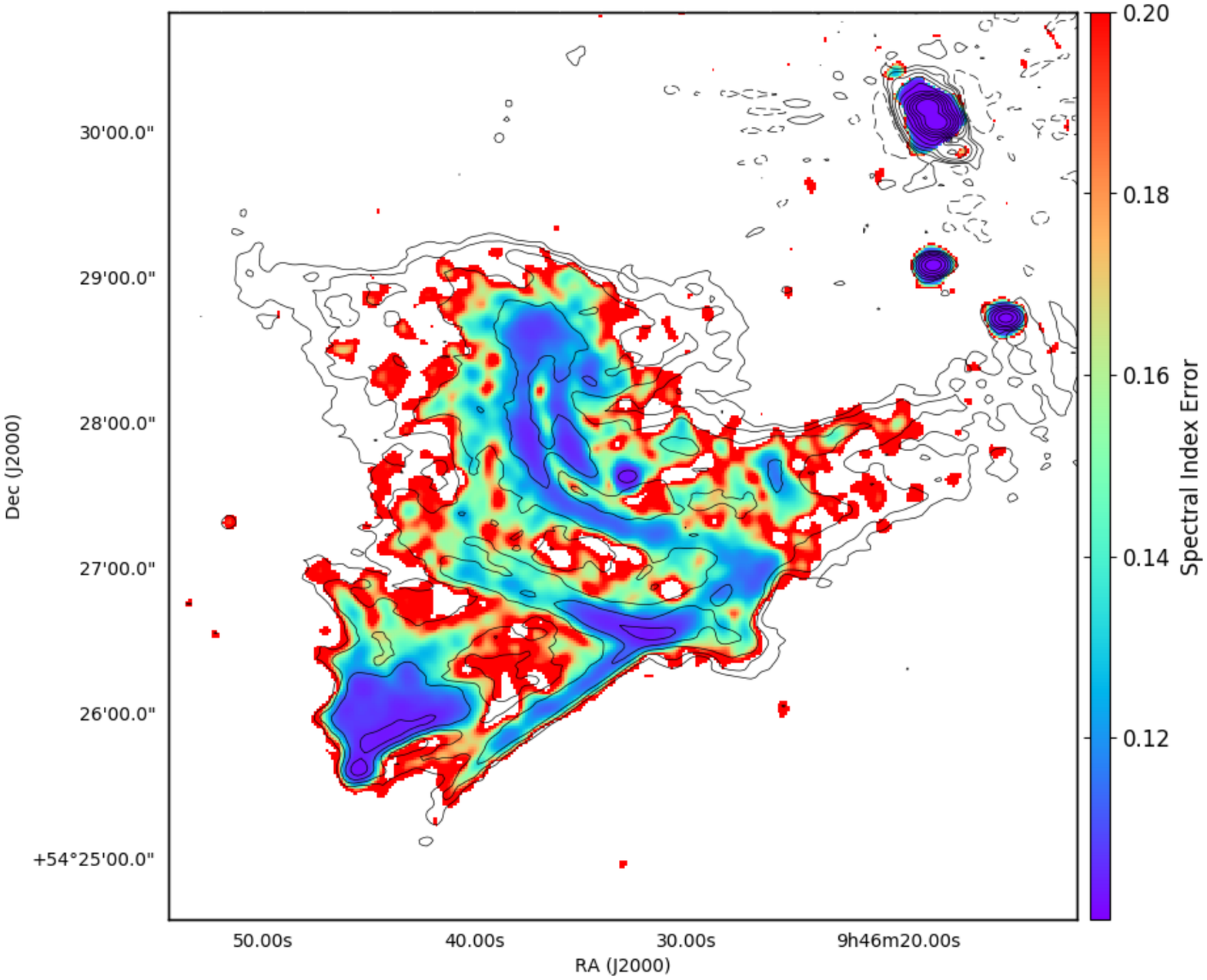}
}
\resizebox{0.49\hsize}{!}{
\includegraphics[angle=0]{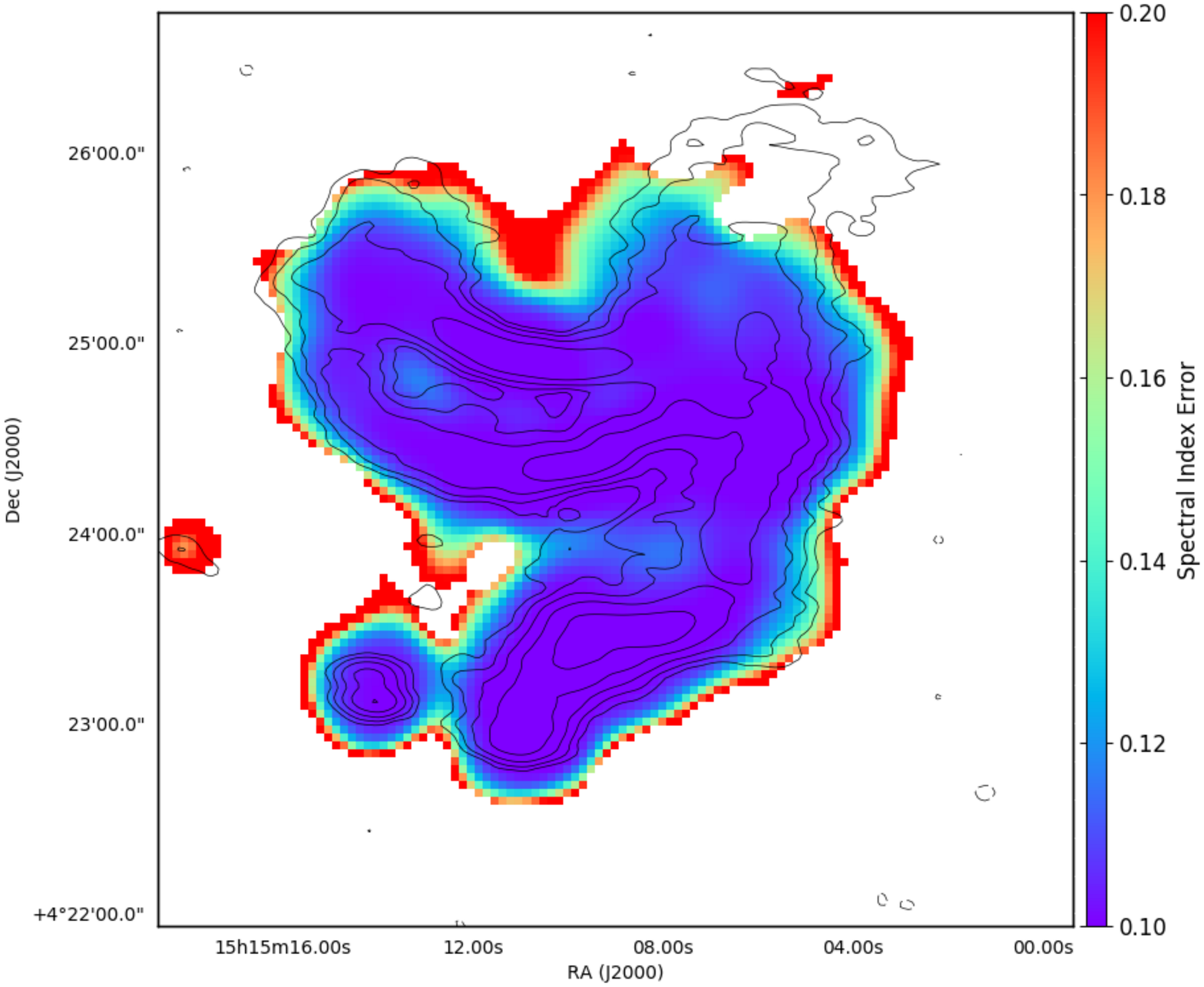}
}
\caption{Spectral index error maps of Abell~2593, SDSS-C4-DR3-3088 and Abell~2048, respectively. The overplotted contours are same as in the LOFAR 150~MHz images for each sources.
}
\label{fig:errormaps}
\end{center}
\end{figure*} 

%\clearpage
\section{Additional spectral index and error maps of SDSS-C4-DR3-3088}

\begin{figure*}[!htb]
\begin{center}
\resizebox{0.49\hsize}{!}{
\includegraphics[angle=0]{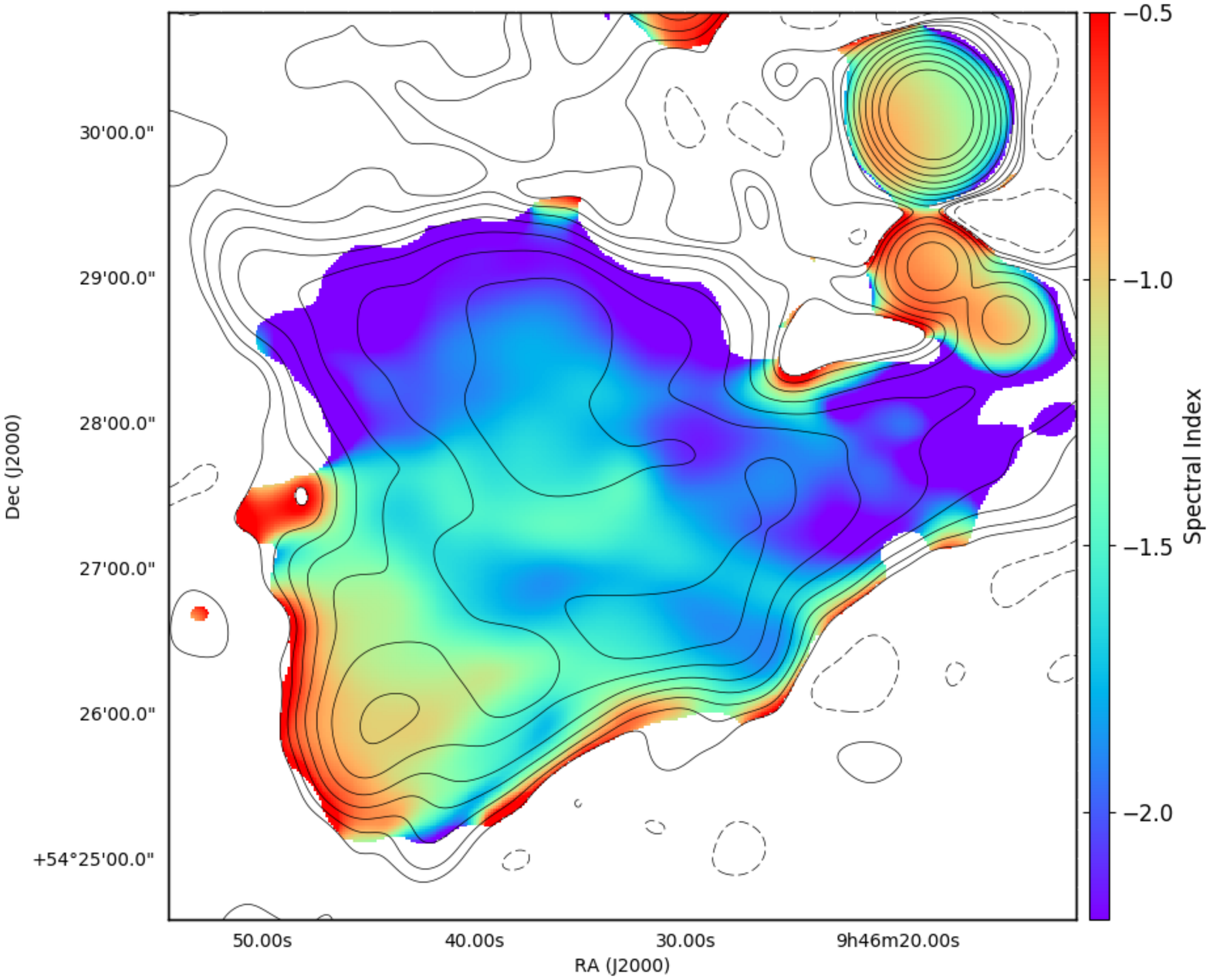}
}
\resizebox{0.49\hsize}{!}{
\includegraphics[angle=0]{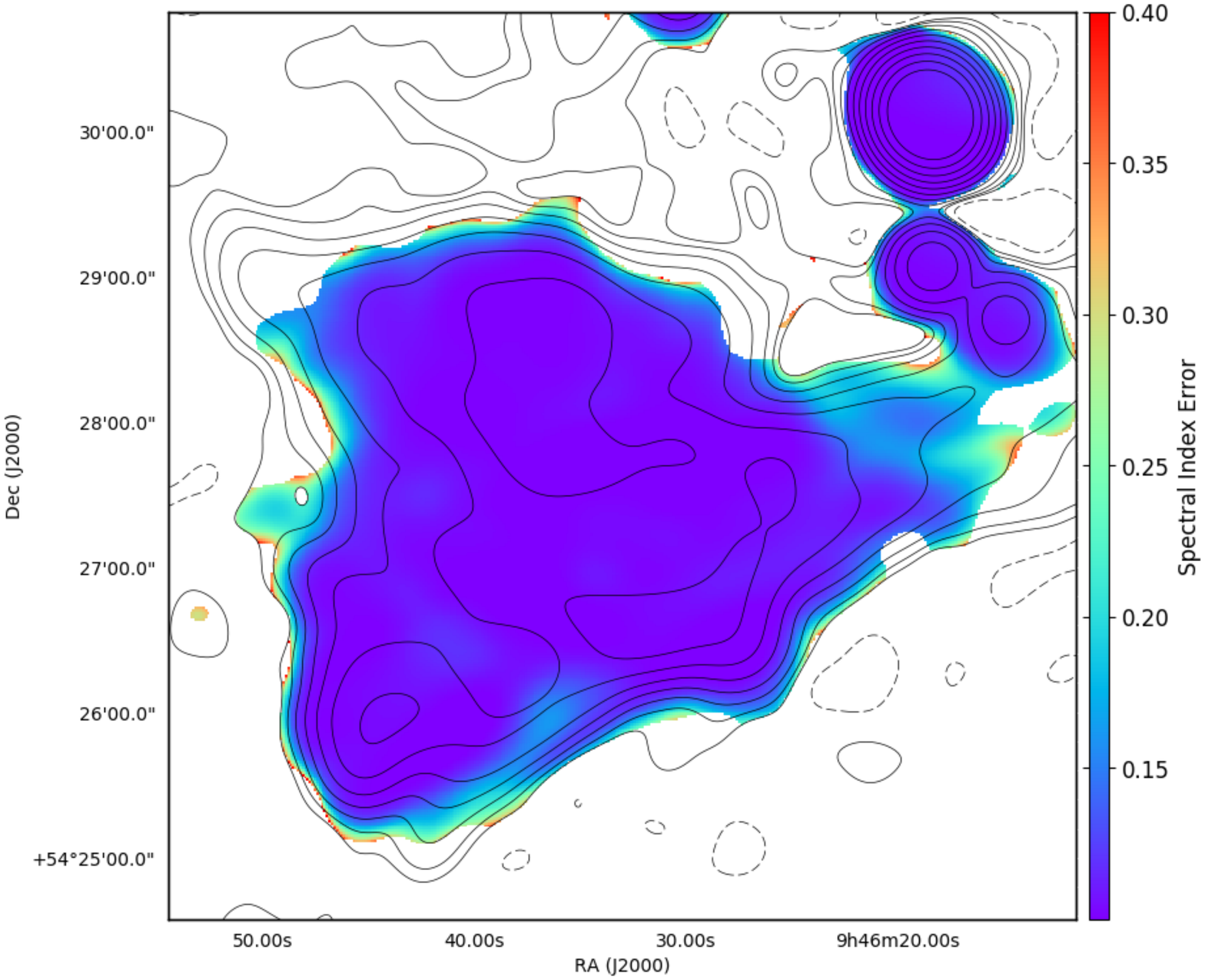}
}
\caption{Low-resolution (\asec{22}) spectral index and error maps of SDSS-C4-DR3-3088.
}
\label{fig:errormaps}
\end{center}
\end{figure*}

\end{appendix}

%\bsp	% typesetting comment
\label{lastpage}
\end{document}